%% file: comp_draft.tex
\documentstyle[times,graphics,epsfig,amsmath,amssymb,natbib,psfrag,color,xspace,ulem,xcolor]{mn2e}

\newcommand{\be}{\begin{equation}}
\newcommand{\e}{\end{equation}}
\newcommand{\bear}{\begin{eqnarray}}
\newcommand{\ear}{\end{eqnarray}}

\def\ctwo{\textsc{C}$^2$\textsc{-ray}\xspace}
\def\cthree{\textsc{CubeP}$^3$\textsc{M}\xspace}
\def\Msol{M_{\odot}}

\def\xh1{x_{ {\rm H~{\sc i}}\,}}
\def\xi{x^{i}_{{\rm H~{\sc{i}}}}\,}
\def\xb{\bar{x}_{{\rm H~{\sc{i}}}}}
\def\Ph1{P_{{H~{\sc{i}}}}}
\def\eh1{\eta_{{H~{\sc{i}}}}}

\def\aap{AAP}
\def\apj{ApJ}
\def\aj{AJ}
\def\apjs{ApJS}
\def\apjl{ApJL}
\def\mnras{MNRAS}

\def\na{New Astronomy}

\def\pasa{Pub. Astro. Soc. Australia}

\def\HI{H~{\sc i}\,}
\def\HII{H~{\sc ii}\,}

\begin{document}

\title[On using semi-numerical simulations of EoR]{On the use of
  semi-numerical simulations in predicting the 21-cm signal from the
  epoch of reionization}

\author[Majumdar et al.]{Suman Majumdar$^{1}$\thanks{E-mail:
    smaju@astro.su.se}, Garrelt Mellema$^{1}$\thanks{E-mail:
    garrelt@astro.su.se}, Kanan K. Datta$^{2}$, Hannes Jensen$^{1}$,
  \newauthor T. Roy Choudhury$^{2}$, Somnath Bharadwaj$^{3}$ and
  Martina M. Friedrich$^{4}$ \\
  $^1$Department of Astronomy \& Oskar Klein Centre, AlbaNova,
  Stockholm University, SE-106 91 Stockholm, Sweden \\ $^2$National
  Centre for Radio Astrophysics, Tata Institute of Fundamental
  Research, Pune - 411007, India\\ $^3$Department of Physics and
  Centre for Theoretical Studies, Indian Institute of Technology,
  Kharagpur - 721302, India\\ $^4$Centro de Ciencias de la
  Atm\'osfera, Universidad Nacional Aut\'onoma de M\'exico, M\'exico}

\date{Accepted 2014 June 29.  Received 2014 June 20; in original form
  2014 March 4}
\maketitle

\begin{abstract}
\input{abs}
\end{abstract}
\input{intro}
\input{sim}
\input{history}

\input{morph}

\input{signal}
\input{summary}

\section*{Acknowledgments} 
The authors would like to thank Ilian T. Iliev for his constructive
comments on the initial version of the manuscript. SM would like to
thank Andrei Mesinger for the useful discussions during a workshop
titled ``Lyman-alpha as an astrophysical tool'' organized by NORDITA
and the Department of Astronomy of Stockholm University in September
2013. KKD thanks the Department of Science \& Technology (DST), India
for the research grant SR/FTP/PS-119/2012 under the Fast Track Scheme
for Young Scientist.

\end{document}

%% file: abs.tex
We present a detailed comparison of three different simulations of the
epoch of reionization (EoR). The radiative transfer simulation (\ctwo)
among them is our benchmark. Radiative transfer codes can produce
realistic results, but are computationally expensive. We compare it
with two semi-numerical techniques: one using the same halos as \ctwo
as its sources (Sem-Num), and one using a conditional Press-Schechter
scheme (CPS+GS). These are vastly more computationally efficient than
\ctwo, but use more simplistic physical assumptions. We evaluate these
simulations in terms of their ability to reproduce the history and
morphology of reionization. We find that both Sem-Num and CPS+GS can
produce an ionization history and morphology that is very close to
\ctwo, with Sem-Num performing slightly better compared to CPS+GS.

We also study different redshift space observables of the 21-cm signal
from EoR: the variance, power spectrum and its various angular
multipole moments. We find that both semi-numerical models perform
reasonably well in predicting these observables at length scales
relevant for present and future experiments. However, Sem-Num performs
slightly better than CPS+GS in producing the reionization history,
which is necessary for interpreting the future observations. The
CPS+GS scheme, however, has the advantage that it is not restricted by
the mass resolution of the dark matter density field.

%% file: intro.tex
\section{Introduction}
\label{sec:intro}
The period in the history of the Universe during which the first
sources of light were formed and the ionizing radiation from these
objects gradually changed the state of hydrogen in the inter-galactic
medium (IGM) from neutral (\HI) to ionized (\HII), is known as the
epoch of reionization (EoR). Our knowledge regarding this epoch is
currently very limited. Observations of the cosmic microwave background (CMBR)
\citep{komatsu11,planck13} and absorption spectra of high redshift
quasars \citep{becker01,fan03,white03,goto11} suggest that
this era probably extended over the redshift range $6 \leq z \leq 15$
\citep{alvarez06,mitra12}. However, these observations are limited in
their ability to shed light on many unresolved but important issues
regarding the EoR, such as the precise duration and timing of
reionization, the relative contributions from various kinds of
sources, the properties of the major sources of ionization, the
typical size and distribution of the ionized regions, etc.

Observations of the redshifted 21-cm signal from neutral hydrogen hold
the possibility to address many of these issues. The brightness
temperature of the redshifted 21-cm radiation is proportional to the
\HI density and is thus in principle capable of probing the \HI
distribution at the epoch where the radiation originated. This
provides a unique possibility for tracking the entire reionization
history. Motivated by this fact several low frequency radio
interferometers such as
{GMRT\footnote{http://www.gmrt.ncra.tifr.res.in}} \citep{paciga13},
{LOFAR\footnote{http://www.lofar.org/}} \citep{yatawatta13,haarlem13},
{MWA\footnote{http://www.haystack.mit.edu/ast/arrays/mwa/}
  \citep{tingay13,bowman13}},
{PAPER\footnote{http://eor.berkeley.edu/}} \citep{parsons13} and
{21CMA\footnote{http://21cma.bao.ac.cn/}} have already started
devoting a significant amount of their observation time towards the
detection of this signal. The future
{SKA\footnote{http://www.skatelescope.org/} \citep{mellema13}} also
has the detection of EoR 21-cm signal as one of its major scientific
goals. However, our lack of knowledge about the properties of the
ionizing sources and different physical processes involved during this
era makes the forecast and interpretation of the expected signal and
the interpretation of the observations of the redshifted 21-cm
radiation very challenging.

A considerable amount of effort has already been devoted to simulate
the expected EoR 21-cm signal. However, the major challenge in such
modelling is the large number of unknown parameters involved and the
huge dynamic range in terms of length scale and mass that one has to
take into account. An accurate model of the EoR should in principle be
able to follow the evolution of the dark matter, gas, radiation and
ionizing sources along with various kinds of possible feedback
processes involved. These simulations need to resolve the low-mass
sources (dark matter halos of mass $\sim 10^8 - 10^9\,{\rm
  M_{\odot}}$) that are expected to dominate the reionization
process. At the same time, simulation boxes need to be large enough
($\sim$ Gpc) to statistically sample the \HI distribution at
cosmological length scales and also to mimic the ongoing and future
\HI survey volumes. Numerical radiative transfer simulations which use
ray-tracing to follow the propagation of ionization fronts in the IGM
are capable of incorporating the detailed physical processes that are
active during reionization
\citep{gnedin00,ciardi01,ricotti02,razoumov02,maselli03,sokasian03,mellema06a,mcquinn07,trac07,semelin07,thomas09}.
Recently, some of these simulations (e.g.\ \citealt{iliev14}) have
been able to achieve the large dynamic range required to do statistics
of the signal on scales comparable to the surveys. However, these
simulations are computationally extremely expensive and it is
therefore difficult to re-run the simulations using different values
for the various mostly unknown reionization parameters.

A computationally much less expensive way of simulating the EoR 21-cm
signal is provided by so-called semi-numerical simulations.  These do
not perform detailed radiative transfer calculations but rather
consider the local average photon density \citep{furlanetto04,
  mesinger07,zahn07, geil08a, lidz09, choudhury09, alvarez09,
  santos10}. In addition to the conventional semi-numerical approach,
recently \citet{battaglia13} have proposed an alternative method to
simulate the 21-cm signal from the EoR, based on the bias between the
underlying density field and the redshift of reionization. Although
using somewhat different approaches, all of these different
semi-numerical simulations are capable of simulating significantly
large volumes of the Universe at reasonably low computational costs.

However, the approximations considered in these semi-numerical schemes
may limit their ability to predict the redshifted 21-cm signal
accurately. To address this issue, \citet{zahn11} performed a
comparison between a set of semi-numerical and radiative transfer
simulations of reionization, using the morphology of the resulting
ionization maps and the spherically averaged real space \HI power
spectrum as metrics for the comparison. Their analysis shows that the
prediction of the real space \HI power spectrum using semi-numerical
schemes differ from the corresponding radiative transfer simulations
by less than $50\%$ during most of the EoR at the length scales of
interest for the present and future surveys.

\citet{zahn11} did not take into account the non-random distortions of
the redshifted 21-cm signal due to peculiar velocities in their
comparison. These so-called redshift space distortions play an
important role in shaping the redshifted 21-cm signal and will
introduce an anisotropy in the 3D power spectrum of the EoR 21-cm
signal
\citep{bharadwaj01,bharadwaj04,barkana05,mao12,shapiro13,majumdar13,jensen13},
similar to the characteristic anisotropy present in the galaxy power
spectrum \citep{kaiser87}. \citet{mesinger11} did compare the
predictions of two semi-numerical schemes with a radiative transfer
simulation using the redshift space 3D spherically averaged \HI power
spectrum. However they included the effect of redshift space
distortions in these simulations in an approximate, perturbative
fashion (similar to \citealt{santos10}), which itself may introduce an
additional error of $\ge 20\%$ in the redshift space \HI power
spectrum \citep{mao12}.

In this paper we present a more thorough and rigorous comparison
between the simulated \HI 21-cm signal generated by a set of
semi-numerical simulations and a radiative transfer simulation. Our
comparison is threefold in nature: we compare these simulations in
terms of their ability to reproduce the reionization history, the
topology of the ionization field at different stages of EoR and
various observable quantities of interest for a redshifted 21-cm
survey to probe the EoR. We implement the redshift space distortions
in our simulation in a more accurate manner than \citet{santos10} and
\citet{mesinger11} by using the actual peculiar velocity fields. The
observable quantities in redshift space that we focus on in this
comparison are the variance of the brightness temperature
fluctuations, the spherically averaged \HI power spectrum and the
ratios of various angular multipole moments of the \HI power spectrum,
which quantify the anisotropies in the signal due to redshift space
distortions. Complementary to the variance of the brightness
temperature fluctuations and the spherically averaged \HI power
spectrum, the angular multipole moments of the \HI power spectrum in
redshift space are expected to provide more information on the history
as well as the intrinsic nature of the reionization
\citep{majumdar13}.

In this work we compare two semi-numerical simulations with a
radiative transfer simulation for hydrogen reionization. The case we
compare is a simplified version of the reionization process in which
we do not include several physical effects whose influence on
reionization are currently not well established. We will mention these
simplifications when we describe our numerical methods. In the
comparison we address the following questions:
\begin{itemize} 

  {\item How well and on which length scales can the semi-numerical
    schemes reproduce the reionization history of a radiative
    transfer simulation?}

  {\item How accurate are the morphologies of the ionization maps that
    are generated by these semi-numerical simulations?}

  {\item How important is it to take into account the effect of 
    redshift space distortions accurately while generating the 21-cm
    signal using these semi-numerical methods?}

  {\item How accurately can different observables of the redshifted 21-cm
    signal (such as the variance, the spherically averaged power
    spectrum and the angular multipole moments of the power spectrum)
    be reproduced by these semi-numerical simulations?}

  {\item Among the two semi-numerical methods discussed here, which
    one is the best considering its capability of generating the
    reionization history, morphology of the ionization maps and the
    observables of the 21-cm signal in redshift space and why?}

\end{itemize}

Throughout this paper we present results for the cosmological
parameters from WMAP five year data release $h= 0.7$, $\Omega_m =
0.27$, $\Omega_{\Lambda} = 0.73$, $\Omega_b h^2 = 0.0226$
\citep{komatsu09}.

The structure of the paper is as follows. In Section 2, we briefly
describe the simulations used. In Section 3, we compare the
reionization history as found by different simulations. Section 4
considers the morphology of the ionization fields generated by different
simulations. In Section 5, we investigate the observable quantities as
predicted by different simulations for a hypothetical redshifted
21-cm \HI survey. We discuss our results and conclude in Section 6.

%% file: sim.tex
\begin{figure}
\begin{centering}
\psfrag{c2ray}[r][r][1][0]{{\ctwo}} 
\psfrag{sem-num e=0.0}[r][r][1][0]{{Sem-Num}} 
\psfrag{21cmFASTL}[r][r][1][0]{{CPS+GS}} 
\psfrag{CPS-NG}[r][r][1][0]{{CPS}} 
\psfrag{xh1}[c][c][1][0]{\large {$\langle \xh1 \rangle_m$}}
\psfrag{z}[c][c][1][0]{\large{$z$}}
\psfrag{xh1vbym}[c][c][1][0]{\large{$\langle \xh1 \rangle_v/\langle
    \xh1 \rangle_m$}}
\includegraphics[width=.4\textwidth,angle=0]{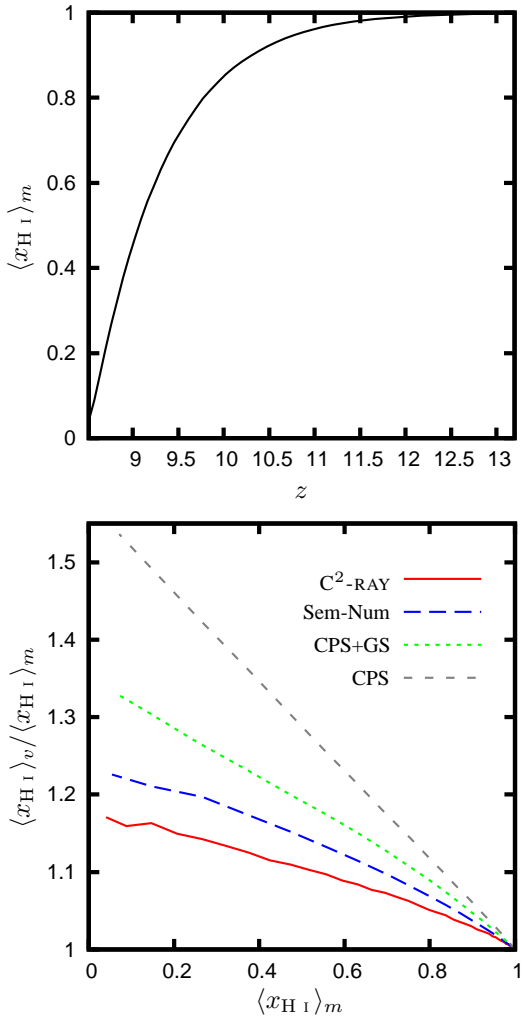}

\caption{The top panel shows the reionization history through mass
  averaged neutral fraction for the radiative transfer
  simulation. Corresponding semi-numerical simulations are tuned to follow
  the same reionization history. The bottom panel shows the evolution
  of the ratio $\langle \xh1 \rangle_v/\langle \xh1 \rangle_m$ for all
  simulations of reionization considered here.}
\label{fig:xh1_his}
\end{centering}
\end{figure}

\section{Simulations}
\label{sec:simulation}
\subsection{$N$-body simulations}
\label{sec:nbody}
All of the reionization simulations discussed in this paper are based
on a single $N$-body run, carried out using the \cthree code
\citep{harnois13}, which is based on \textsc{PMFast}
\citep{merz05}. \cthree uses a particle-particle-particle-mesh scheme,
calculating short-range gravitational interactions directly between
particles and long-range interactions on a grid. For performance
reasons, this grid is split into a fine local grid and a coarser
global grid. For the simulations considered here, we used a simulation
volume of ($163$ Mpc)$^3$ (comoving) with a fine grid consisting of
$6144^3$ cells. The number of $N$-body particles was $3072^3$.

For each output from the $N$-body simulations, halos were identified
using a spherical over-density method. This method encloses local
density maxima in progressively larger spheres until the average
density of the sphere goes below $178$ times the global mean density. We
allowed halos down to $20$ particles, corresponding to $10^8
\;\Msol$. After constructing the halo lists, the $N$-body particles
for each output were smoothed onto a grid with $256^3$ cells to
produce the density field. This $N$-body simulation was described in
more detail in \citep{iliev12}.

\subsection{Radiative Transfer  simulations}
\label{sec:RT}
For the radiative transfer simulations, we used \ctwo
\citep{mellema06b}---``Conservative Causal Ray-tracing method''. \ctwo
works by tracing rays from all sources and iteratively solving the
equation for the time evolution of the ionization fraction of hydrogen
($x_i$) as,
\begin{equation}
  \frac{d x_i}{d t} = (1 - x_i) (\Gamma + n_e C_H) - x_i n_e
  C\alpha_{\rm B},
  \label{eq:ionz_ev}
\end{equation}
where $\Gamma$ is ionization rate, $n_e$ is the density of free
electrons, $C_H$ is the collisional ionization coefficient, $C$ is the
so-called clumping factor and $\alpha_{\rm B}$ is the recombination
coefficient. The clumping factor $C$ is defined as $\langle
n^2\rangle/\langle n\rangle^2$, where the average is taken over the
volume resolution of the density field, in our case
$163/256=0.64$~Mpc. This factor takes into account the effect of
density variations below the resolution scale. This clumping factor
depends on the density variations in the gas which will be both time-
and position-dependent. There exist several approximate recipes to
include the effect of subgrid clumping. However, since there is no
consensus on the best way we take $C$ to be 1 in our comparison,
although we note that in reality it is expected to be larger than
this.

Eq. (\ref{eq:ionz_ev}) is solved by iterating over each
cell and each source until convergence. By using the time-averaged
$\Gamma$ for each time step, \ctwo is able to use relatively large
time-steps while still conserving photons (see \citealt{mellema06b}).

In principle \ctwo can incorporate various kinds of source model. For
this work, the sources from the $N$-body simulations described above
were assigned ionizing fluxes $\dot{N}_{\gamma}$ proportional to the
halo mass $M_{\mathrm{h}}$ as,
\begin{equation}
  \dot{N}_{\gamma} = g_{\gamma} \frac{M_{{\rm h}} \Omega_{{\rm b}}}{(10\;{\rm Myr}) \Omega_{{\rm m}} m_{{\rm p}}},
  \label{eq:ionz_flux}
\end{equation}
where $m_{\mathrm{p}}$ is the proton mass and $g_{\gamma}$ is a source
efficiency coefficient, which in effect depends on the star formation
efficiency, the initial mass function and the escape fraction. In this
particular simulation we have assumed that only those sources
contribute to reionization which have mass $\ge 2.2 \times 10^9 \Msol$
and we set $g_{\gamma} = 21.7$ for all of them. This simulation was
previously described in \citet{iliev12} as ``L3''. We selected this
simulation since it does not use any suppression of sources, a process
which is not included in the semi-numerical methods we use here. This
is another simplification we introduce in order to compare the
radiative transfer and semi-numerical methods at the more basic level
without the introduction of subgrid physics whose influence on
reionization is currently not well established.
 
The reionization history for this model is illustrated through the
evolution of the mass averaged neutral fraction ($\langle \xh1
\rangle_m$, also represented by $\xb$ for convenience in the rest of
the paper) in the top panel of Figure \ref{fig:xh1_his}.

\subsection{Semi-numerical simulations}
\label{sec:SN}
\ctwo and similar kinds of radiative transfer algorithms are capable
of generating an accurate reionization topology and history since they
take into account the ionization and recombination processes
(eq. [\ref{eq:ionz_ev}]) along the path of each individual
photon. However, to achieve this level of precision they require huge
amounts of computational time (hundreds of thousands of core
hours). Thus, it would be very expensive to explore the mostly unknown
parameter space of possible reionization models using this kind of
simulations. Furthermore, most of the present and upcoming radio
interferometric reionization surveys (including the humongous SKA)
will not be sensitive enough to map the \HI distribution from this
epoch with a precision and resolution comparable to that of the
simulations. These limitations of the radiative transfer simulations
as well as the poor sensitivity of the present and future EoR 21-cm
surveys together have motivated the development of approximate
semi-numerical methods to simulate the redshifted 21-cm signal from
EoR. These approximate methods are expected to simulate the \HI 21-cm
signal from this epoch accurately enough for the length scales to
which the present and upcoming 21-cm surveys will be sensitive, at a
very nominal computational cost. They can simulate a reasonable volume
of the universe (comparable to the survey volume of LOFAR or SKA) in a
few minutes of computational time on a single processor with
considerably less memory consumption (few gigabyte of RAM). If
semi-numerical simulations are accurate enough in predicting the
redshifted 21-cm signal from EoR, one can achieve almost a five orders
of magnitude gain in computational time compared to a radiative
transfer simulation.

Most of the conventional semi-numerical methods of simulating EoR are
based on comparing the average number of photons in a specific volume
with the average number of neutral hydrogen present in that
volume. This is the basic principle of the excursion-set formalism
developed by \citet{furlanetto04}. We discuss two different approaches
of implementing it in the following sections. One important common
feature of the two semi-numerical simulations discussed here is that
the ionization map generated by them at each redshift is dependent
only on the matter distribution or the matter and source distribution
at that specific redshift. Due to this it is possible to generate the
ionization maps at several redshifts simultaneously (or in parallel)
using these simulations.

\subsubsection{Semi-numerical simulation with halos (Sem-Num)}
\label{sec:Sem-Num}
The first of the semi-numerical methods we use here is based on the
excursion-set formalism of \citet{furlanetto04} and similar to
\citet{zahn07}, \citet{mesinger07}, \citet{choudhury09} and
\citet{santos10}. Here we assume that the halos are the sites where
the ionizing photon emitting sources were formed. To date, little is
known about the high redshift photon sources and the characteristics
of their radiation, so this method assumes that the total number of
ionizing photons contributed by a halo of mass $M_{\rm h}$ is
\begin{equation}
N_{\gamma}(M_{\rm h}) = N_{\rm ion} \frac{M_{\rm h} \Omega_{{\rm
      b}}}{m_{\rm p}\Omega_{{\rm m}}} 
\label{eq:sem-num}
\end{equation}
where $N_{\rm ion}$ is a dimensionless constant, which effectively
represents the number of photons entering in the IGM per baryon in
collapsed objects. In this paper we have assumed that
$N_{\gamma}(M_{\rm h})$ is proportional to the halo mass $M_{\rm h}$
but in principle one can assume any functional form for
$N_{\gamma}(M_{\rm h})$. This particular source model
(eq. [\ref{eq:sem-num}]) is thus similar to that of \ctwo
(eq. [\ref{eq:ionz_flux}]). The assumptions regarding the source model
play a crucial role in the resulting ionization and brightness
temperature fields from a semi-numerical simulation, as will become
more clear in the next few sections.

Once the locations and masses of the halos are known and a functional
form for $N_{\gamma}(M_{\rm h})$ has been assigned, the ionizing
photon field can be constructed. To construct the ionization field, we
estimate the average number density of photons $\langle n_{\gamma}
({\bf x})\rangle_R$ within a spherical region of radius $R$ around a
point ${\bf x}$ and compare it to the corresponding spherically
averaged number density of neutral hydrogen $\langle n_{\rm H}
\rangle_R$. The radius of this smoothing region is then gradually
increased, starting from the grid cell size ($R_{\rm cell}$) and going
up to a certain $R_{\rm max}$, which is determined by the assumed mean
free path of the photon at the concerned redshift. We consider the
point ${\bf x}$ to be {ionized\footnote{The main difference between
    \citet{mesinger07} and this approach is the following. In our
    simulation we assume that only the central pixel of the smoothing
    sphere is ionized when the ionization condition is satisfied,
    whereas in \citet{mesinger07} it is assumed that the entire region
    inside the smoothing sphere is ionized. In this sense, our method
    of flagging ionized cells is similar to what is done in
    \citet{mesinger11}.}} if the condition
\begin{equation}
  \langle n_{\gamma} ({\bf x})\rangle_R \ge \langle n_{\rm H}
  \rangle_R (1+ {\bar N}_{\rm rec})
\label{eq:sem-num-ion}
\end{equation}
is satisfied for any smoothing radius $R$, where ${\bar N}_{\rm rec}$
is the average number of {recombinations\footnote{It is also possible
    to incorporate a self-shielding criterion in this simulation based
    on a density dependent recombination scheme (eq. [15] of
    \citealt{choudhury09}), which we do not consider in this work.}}
per hydrogen atom in the IGM. Note that various other unknown
parameters {\it e.g.} star formation efficiency within halos, number
of photons per unit stellar mass, the photon escape fraction, helium
weight fraction, as well as the factor $(1 + {\bar N}_{\rm rec})$ can
be absorbed within the definition of $N_{\rm ion}$ and we do so in
this work. In other words, in this description the effect of
recombinations can be compensated by making the sources more
efficient. It also means that the effect of recombinations is here
taken to be uniform over the whole volume, although in reality it will
be position-dependent. Because of this, we have chosen the clumping
factor in eq. (\ref{eq:ionz_ev}) to be 1, so as to make the treatment
of recombinations more similar between the semi-numerical and
numerical methods.

We apply periodic boundary conditions when calculating the
ionization field. Points where the above ionization condition is not
satisfied, are given an ionization fraction equal to $\langle
n_{\gamma} ({\bf x})\rangle_{R_{\rm cell}}/\langle n_{\rm H}
\rangle_{R_{\rm cell}}$. This approximately takes into account the
\HII regions not resolved by the resolution of the simulation
\citep{geil08a}. Finally, we tune the value of $N_{\rm ion}$ in such a
way that we achieve the same evolutionary history for $\langle \xh1
\rangle_{m}$, as \ctwo (see top panel of Figure
\ref{fig:xh1_his}). Hereafter we refer to this method as ``Sem-Num''.

\subsubsection{Conditional Press-Schechter (CPS and CPS+GS)}
\label{sec:CPS}
The second semi-numerical method we consider here is based on
the conditional Press-Schechter formalism initially proposed by
\citet{bond91} and \citet{lacey93} and later modified by
\citet{barkana04, barkana08}. Unlike the previous semi-numerical
simulation, where the halos are the locations of the ionizing
sources, this method is solely based on the underlying matter density
field. According to this scheme the collapsed fraction at a redshift
$z$ within a region of size $R$ depends on the mean overdensity of
that region ${\bar\delta_R}$ as
\begin{equation}
  f_{\rm coll} = \frac{{\bar f}_{\rm ST}}{{\bar f}_{\rm PS}}\, {\rm erfc} \left[ \frac{\delta_c (z) - {\bar\delta_R}}{\sqrt{2\left[ \sigma^2(R_{\rm min}) - \sigma^2(R) \right]}}\right]\,,
  \label{eq:coll}
\end{equation}
where $R_{\rm min}$ is the radius that encloses the mass $M_{\rm min}$
at average density ${\bar\rho}$, $\delta_c (z)$ is the critical over
density required for spherical collapse and has the redshift
dependence $\delta_c (z) = 1.686/D(z)$, $D(z)$ is the linear growth
factor, $\sigma^2(R)$ is the linear theory rms fluctuation of the
density on scale $R$, ${\bar f}_{\rm ST} (z, R_{\rm min})$ is the mean
Sheth-Tormen collapsed fraction with the normalization of
\citet{jenkins01} and ${\bar f}_{\rm PS} (z, R_{\rm min}, R)$ is the
mean Press-Schechter collapsed fraction estimated from the density
field at redshift $z$ after being smoothed over a length scale of size
$R$. We set $M_{\rm min} = 2.2 \times 10^9 M_{\odot}$ at all
redshifts, to keep it consistent with the minimum halo mass used in
the other simulations of reionization considered in this paper. One
advantage of this simulation method over any halo based simulation
scheme (semi-numerical or radiative transfer) is that it is not
restricted by the particle mass resolution required to identify the
atomically cooling halos ($M_{\rm min} \sim 10^8 M_{\odot}$) or even
molecularly cooling halos ($M_{\rm min} \sim 10^6-10^8
M_{\odot}$). This allows us to include the contribution from
atomically and molecularly cooling halos without detecting them
individually. These sources could contribute substantially to the
reionization process. We note, however, that for a given resolution,
the CPS value for $f_{\rm coll}$ is not identical to the numerical
one.

For a specific smoothing scale $R$, a point is considered to be
ionized if the collapsed fraction calculated for a smoothing region of
size $R$ around it is more than the inverse of the ionizing efficiency
$\zeta$
\begin{equation}
  f_{\rm coll} \ge \zeta^{-1} \,.
\end{equation}
Similar to the $N_{\rm ion}$ in the previous simulation, various
parameters of reionization can be incorporated into $\zeta$. This
simulation model is similar to the models of
\citet{zahn05,alvarez09,zahn11,mesinger11}. In this scheme, by
default, uniform or no {recombination\footnote{Recently
    \citet{sobacchi14} have developed a model to implement a density
    dependent recombination scheme which can be combined with this
    simulation model. We do not consider it in this work.}} is assumed
for every part of the density field. In an earlier work,
\citet{zahn11} have reported that the ionization map shows a better
match with the radiative transfer simulations when the smoothing is
done with a sharp $k$-space filter instead of a spherical top hat
filter. However, we observe that both the spherical top hat and the
sharp $k$-space filter produce very similar ionization maps in this
case (when compared in terms of their bubble size distribution and
power spectrum). In this work we choose to use a sharp $k$-space
filter for the smoothing of the density field as it is expected to
keep the photon number conserved in comparison to the spherical top
hat filter \citep{zahn07}.

Irrespective of what filter we use, this simulation technique tends to
produce a much stronger ``inside-out'' reionization than other two
models (an initial indication of this can be seen in the bottom panel
of Figure \ref{fig:xh1_his}). This leads to the production of more
small scale ionized regions at any stage of reionization. The density
field used in \citet{mesinger11} was constructed at a very high
redshift using Zel'dovich approximations and then extrapolated to
redshift $z$. The matter distribution in such a density field is
expected to be slightly more diffuse (or less clustered) in nature
than the one obtained using an $N$-body simulation, as in our
case. This inherent diffuse nature of the density field probably
prevents the production of a large number of small scale ionized
regions. To generate a similar sort of diffused density field, we
convolve the N-body density field with a Gaussian filter of width
$\approx 2$ Mpc (equivalent to the size of $3$ grid cells in this
case). The collapsed fraction is then estimated from this density
field following eq. (\ref{eq:coll}). As we will see in the following
sections, this treatment makes the output from this simulation more
similar to the other two discussed here. To achieve the same $\langle
\xh1 \rangle_{m}$ evolution as that of the \ctwo, we adjust the value
of $\zeta$. Hereafter we refer to this simulation method as
``CPS+GS''. For most of our comparison analysis in this paper we have
used these Gaussian-smoothed density fields for CPS+GS but we have
also used unsmoothed density fields (hereafter referred to as ``CPS'')
for some test comparisons.

In Table \ref{tab:sim}, we briefly summarize the major characteristics
of all the simulations discussed here.
\begin{table}
\centering
\begin{tabular}{c|c|c|c}
\hline
\hline
 Simulation & Density field & Halos used & Ionization field\\
  & used & as sources & obtained by \\
\hline
 \ctwo & $N$-body & Yes & Radiative transfer\\
 Sem-Num & $N$-body & Yes & Excursion-set based\\
 CPS & $N$-body & No & Excursion-set based\\
 CPS+GS & $N$-body+GS & No & Excursion-set based\\
\hline
\hline
\end{tabular}
\caption{ The major characteristics of the different
  simulations considered here.}
\label{tab:sim}
\end{table}

%% file: history.tex
\begin{figure*}
\psfrag{C2RAY}[c][c][1][0]{\large{\ctwo}}
\psfrag{Sem-Num}[c][c][1][0]{\large{Sem-Num}}
\psfrag{CPS}[c][c][1][0]{\large{CPS+GS}}
\psfrag{Mpc}[c][c][1][0]{\large{Mpc}}
\psfrag{z}[c][c][1][0]{\large{$z$}}
\psfrag{xi}[c][c][1][0]{\large{$x_{{\rm i}}$}}
\includegraphics[width=1.\textwidth,
  angle=0]{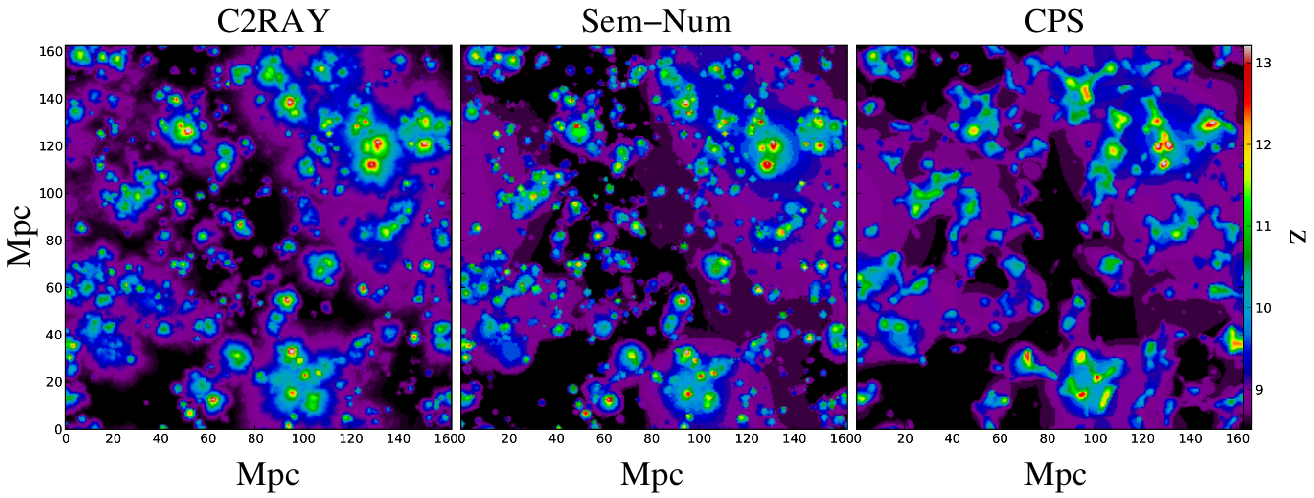}
\includegraphics[width=1.\textwidth,
  angle=0]{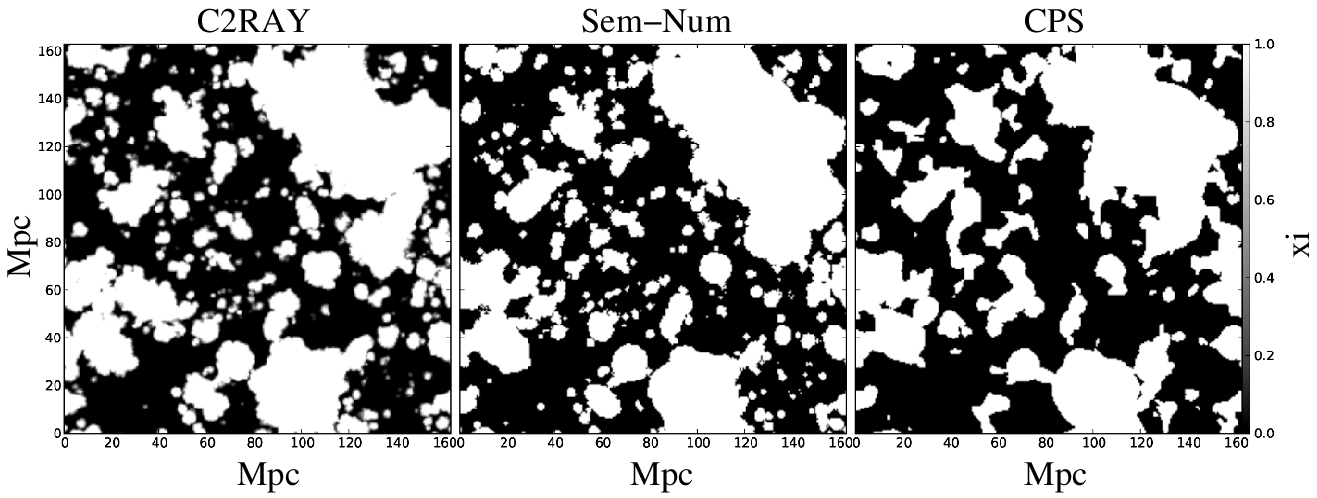}
  \caption{\textit{Top panel:} the redshift of
    reionization in each individual cell for three different
	simulations. \textit{Bottom panel:} the state of ionization in
    three different simulations when the reionization was half way
    through ($z=9.026$). The thickness of each slice is $0.64$ Mpc.}
\label{fig:his_map}
\end{figure*}

\section{Reconstruction of the reionization history}
\label{sec:history}
\begin{figure*}
\psfrag{C2RAY}[r][r][1][0]{{\ctwo}}
\psfrag{Sem-Num e=0.0}[r][r][1][0]{{ Sem-Num }}
\psfrag{21cmFASTL}[r][r][1][0]{{ CPS+GS }}
\psfrag{CPS-NG}[r][r][1][0]{{ CPS }}
\psfrag{bias}[c][c][1][0]{\large{$b_{z\Delta}(k)$}}
\psfrag{r}[c][c][1][0]{\large{$r_{z \Delta}(k)$}}
\psfrag{rcross}[c][c][1][0]{\large{${\mathcal R}_{zz}(k)$}}
\psfrag{k (Mpc)}[c][c][1][0]{\large{$k\, ({\rm Mpc}^{-1})$}}
\includegraphics[width=1.0\textwidth,angle=0]{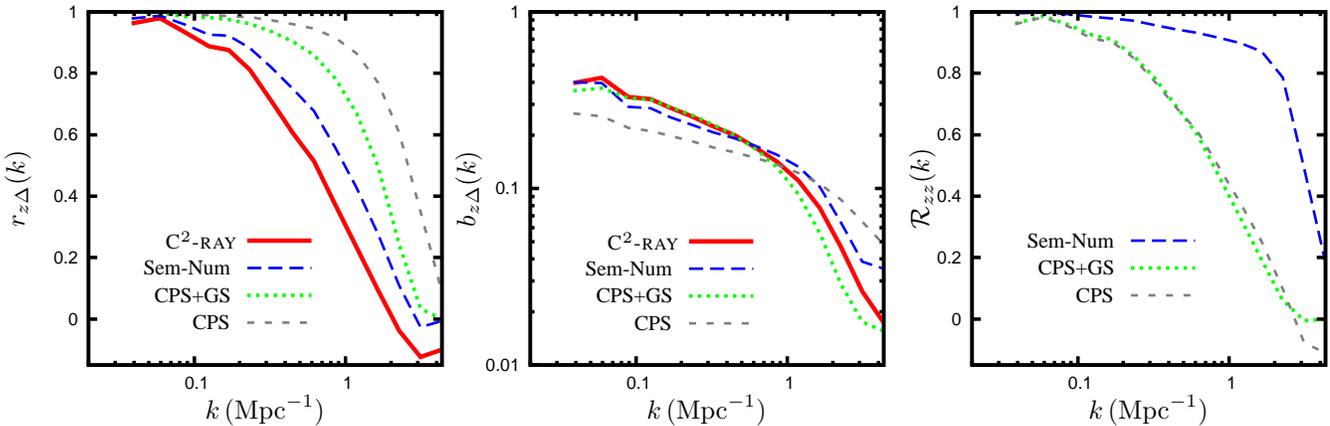}
\caption{First two panels from left show the cross-correlation $r_{z
    \Delta}(k)$ between the density field at the mid-point of
  reionization ($z=9.026$) and the redshift of reionization field and
  the bias $b_{z\Delta}(k)=\sqrt{P_{zz}(k)/P_{\Delta \Delta}(k)}$
  between the same quantities for different simulations,
  respectively. The third right most panel shows the cross-correlation
  ${\mathcal R}_{zz}(k)$ between the redshift of reionization fields
  for \ctwo and the same from different semi-numerical simulations.}
\label{fig:his_bias_corr}
\end{figure*}
As mentioned in the previous section, the ionization field produced by
these two semi-numerical simulations at a specific redshift will not
have any memory of the ionization field at an earlier
redshift. However, it is possible to chronologically follow the
reionization history in these semi-numerical simulations by
sequentially producing ionization maps using the previously produced,
higher-redshift, maps as input. This will, however, slow down the
semi-numerical schemes by not allowing them to generate the ionization
maps at a number of redshifts simultaneously (or in parallel).

In this work, we have run our semi-numerical schemes following the
usual convention ({\it i.e.} we have not followed the ionization state
of each grid point chronologically to determine its ionization state
at a later redshift).  Generally, the values of $N_{\rm ion}$ or
$\zeta$ are adjusted in such a way that these simulations
approximately produce the same $\langle \xh1 \rangle_{m}$ or $\langle
\xh1 \rangle_{v}$ evolution as found in radiative transfer
simulations. We adopt the same approach in this work. However, to
interpret the EoR redshifted 21-cm signal from future surveys with
these approximate simulations, they should be capable of reproducing
the reionization history with a certain acceptable level of
accuracy. In this section we explore up to what extent they can
reliably mimic the reionization history.

We first study the evolution of the volume-averaged neutral fraction
produced by different simulations. As mentioned earlier, both of the
semi-numerical simulations are tuned to produce the same evolution for
the mass-averaged neutral fraction ($\langle \xh1 \rangle_{m}$) as
that of the \ctwo. However, due to the differences in their approach,
the evolution of $\langle \xh1 \rangle_{v}$ is not necessarily the
same for the four simulations. The bottom panel of Figure
\ref{fig:xh1_his} shows the ratio $\langle \xh1 \rangle_{v}/\langle
\xh1 \rangle_{m}$ as a function of $\langle \xh1 \rangle_{m}$ for the
four different simulations discussed earlier. It is clear from this
plot that at almost any stage of reionization (except the very
beginning), $\langle \xh1 \rangle_{v}$ is always smaller for \ctwo
than for the two semi-numerical simulations. This difference gradually
increases as reionization progresses. This implies that less volume is
ionized in case of the two semi-numerical simulations with respect to
\ctwo to achieve the same mass averaged ionization fraction. This
further implies that the ionization maps in semi-numerical simulations
follow the density field more closely than the radiative transfer
simulation (we will elaborate on this point further in the following
sections). Among the two semi-numerical schemes, CPS+GS has a higher
value of $\langle \xh1 \rangle_{v}/\langle \xh1 \rangle_{m}$ than
Sem-Num at any stage and this difference goes up to approximately
$10\%$ at the very late stages of reionization. When the Gaussian
smoothing of the density field is not done in CPS, the difference
between Sem-Num and CPS can go up to approximately $30\%$.

Next, we show how well these semi-numerical simulations are able to
reconstruct the history at the level of each individual grid cell. To
do so, we have saved the redshift of reionization of each grid point
for four different simulations. We have used an ionization threshold
of $x_{th} \ge 0.5$ to identify a cell as ionized. The top panel of
Figure \ref{fig:his_map} shows one slice of the simulation box with a
colour coded map for redshift of reionization for the three different
schemes. A simple visual inspection of this image along with the
ionization state of the same slice at the mid point of reionization
(bottom panel of Figure \ref{fig:his_map}) suggest that the
reionization history reproduced by Sem-Num resembles that of the \ctwo
simulation more than CPS+GS does. The redshift map of CPS+GS looks
smoother than the other two simulations. This is a clear signature of
the more diffuse matter distribution that was used in CPS+GS. We find
that the reionization redshift map for Sem-Num is highly correlated to
that of the \ctwo. The correlation coefficient between these two maps
(right most panel in Figure \ref{fig:his_bias_corr}) has a value $\ge
0.9$ for length scale range $k \leq 1.0 \,{\rm Mpc}^{-1}$. The same
correlation between CPS+GS and \ctwo is $\ge 0.4$ for $k \leq 1.0
\,{\rm Mpc}^{-1}$.

We estimate the bias and cross-correlation between the reionization
redshift and the matter density field to quantify how the reionization
history is related to (or rather controlled by) the underlying matter
distribution. We define the fluctuations in the redshift of
reionization field as $\delta_z ({\bf x}) = \left\{[1+ z({\bf x})] -
  [1+{\bar z}]\right\}/[1+{\bar z}]$ and similarly for a density field
at a specific redshift as $\delta = \left[\rho({\bf x}) - {\bar
    \rho}\right]/{\bar \rho}$, where ${\bar z}$ and ${\bar \rho}$ are
the means of the corresponding reionization redshift and the density
field respectively. Thus the bias and cross-correlation between these
two fields in Fourier space can be defined as $b_{z\Delta}(k) =
\sqrt{P_{zz}(k)/P_{\Delta \Delta}(k)}$ and $r_{z \Delta}(k) = P_{z
  \Delta}(k)/\sqrt{P_{zz}(k)P_{\Delta \Delta}(k)}$, respectively. The
quantities $P_{zz}$ and $P_{\Delta \Delta}$ are the power spectrum of
the field $\delta_z$ and $\delta$ respectively and $P_{z \Delta}$ is
the cross-power spectrum between these two fields.

We calculate the bias factor and the cross-correlation between the
reionization redshift field and the matter density field at
approximately the mid point of reionization ({\it i.e.} at $z=9.026$
when $\langle \xh1 \rangle_{m} \simeq 0.5$). The central panel in
Figure \ref{fig:his_bias_corr} shows the bias $b_{z\Delta}(k)$ for the
four different simulations. In all four, the bias between redshift and
density is highest at the largest scale and gradually decreases at
smaller scales. The bias estimated from Sem-Num and CPS+GS are in very
good agreement with that of \ctwo (less than $5\%$ difference) for $k
\leq 1.0 \,{\rm Mpc}^{-1}$, whereas the bias estimated from CPS (in
which Gaussian smoothing of the density field is not done) is lower
than that of \ctwo ($\sim 30-50\%$ less) in the same length scale
range. We find that the reionization history in the case of CPS+GS is
more strongly correlated with the density field ($r_{z \Delta} \ge
0.8$ for $k \leq 1.0 \,{\rm Mpc}^{-1}$; left most panel in Figure
\ref{fig:his_bias_corr}) compared to \ctwo and Sem-Num ($r_{z \Delta}
\ge 0.3$ and $\ge 0.5$, respectively for the same $k$ range).  The
cross-correlation between same two fields in case of CPS is even
higher ($r_{z \Delta} \ge 0.9$ for $k \leq 1.0 \,{\rm Mpc}^{-1}$).

This statistical analysis suggests that Sem-Num is capable of
producing a more reliable reionization history compared to CPS+GS or
CPS at the length scales corresponding $k \leq 1\,{\rm Mpc}^{-1}$. The
estimated cross-correlation $r_{z \Delta}$ also suggests that for
CPS+GS and CPS reionization is more strongly correlated with the
density field than the other two schemes.

\citet{battaglia13} constructed an empirical model of reionization by
extrapolating the bias $b_{z\Delta}(k)$ and cross-correlation $r_{z
  \Delta}$ of a radiative transfer simulation. We find that even when
the history of reionization is not followed chronologically at the
level of individual grid cells, a semi-numerical simulation like
Sem-Num is still capable of generating the same bias and
cross-correlation as that of the radiative transfer simulation at
length scales relevant for the present and upcoming EoR 21-cm
experiments. The main difference between Sem-Num and CPS+GS or CPS is
in the source model. The source model of Sem-Num
(eq. [\ref{eq:sem-num}]) is very similar to that of the \ctwo
(eq. [\ref{eq:ionz_flux}]). We can thus safely say that a
semi-numerical scheme can reliably reproduce the reionization history
of a radiative transfer simulation when a similar source model is used
in it.

%% file: morph.tex
\section{Comparison of the Morphology of the Ionization Maps}
\label{sec:morpho}
\begin{figure*}
\psfrag{z=10.290, xh1=0.898}[c][c][1][0]{{$\xb=0.90$}}
\psfrag{z=9.164, xh1=0.557}[c][c][1][0]{{$\xb=0.56$}}
\psfrag{z=8.892, xh1=0.375}[c][c][1][0]{{$\xb=0.38$}}
\psfrag{z=8.636, xh1=0.147}[c][c][1][0]{{$\xb=0.15$}}
\psfrag{C2RAY}[r][r][1][0]{{\ctwo}}
\psfrag{Sem-Num e=0.0}[r][r][1][0]{{Sem-Num}}
\psfrag{21cmFASTL}[r][r][1][0]{{CPS+GS}}
\psfrag{CPS-NG}[r][r][1][0]{{CPS}}
\psfrag{R dP/dR}[c][c][1][0]{{\small $R dP/dR$}}
\psfrag{R (Mpc)}[c][c][1][0]{{\small $R\, ({\rm Mpc})$}}
\includegraphics[width=1.\textwidth,
  angle=0]{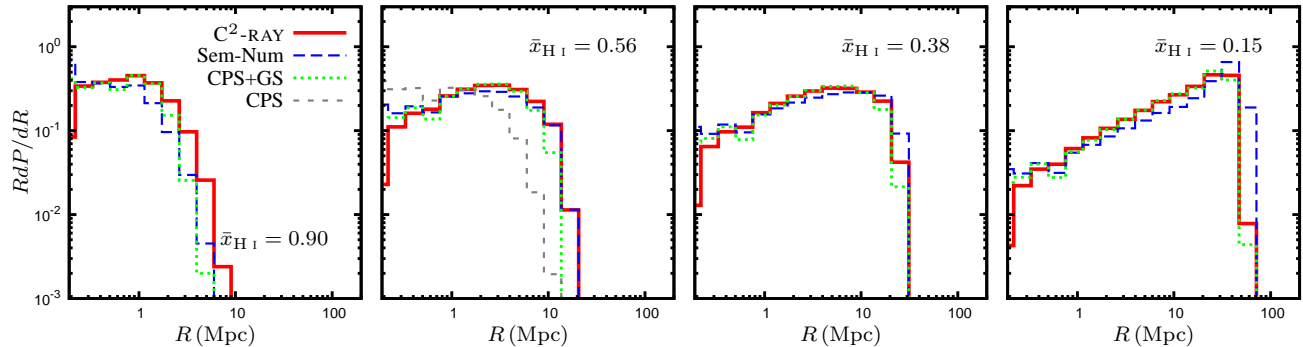}

  \caption{The spherical average bubble size distribution at four
    representative values of $\xb$. We show the distribution for CPS
    at $\xb=0.56$ only, to illustrate its marked difference from
    \ctwo.}
\label{fig:spa_spec}
\end{figure*}
The evolution of the morphology of the ionization field directly
controls the evolution of the redshifted \HI 21-cm signal. In this section  
we use a few different statistical
measures to analyze and quantify the morphology of the ionization maps
generated by the three different simulation methods. Some of these
morphological similarities can be seen easily from a
rough visual inspection of the ionization maps (see bottom panels of
Figure \ref{fig:his_map} and {brightness
  temperature\footnote{The brightness temperature is directly proportional
    to the neutral fraction, so these brightness temperature maps have
    a one-to-one correspondence with the ionization maps. Note that
    the brightness temperature maps shown here are in redshift
    space.}} maps of Figure \ref{fig:dt_maps}). However, such a visual
inspection will also reveal some of their differences. In all
three simulation methods discussed here, the ionized regions
essentially follow the distribution of the ionizing sources ({\it
  i.e.}  the distribution of the halos or the high density peaks) at
the very early stages of reionization and are small in size. As the
time progresses, the \HII regions gradually get larger in size and
start merging with each other. At the very late stages of reionization
the ionization fronts start progressing into the low density regions
and finally, almost the entire IGM is ionized.

Among the two semi-numerical schemes, Sem-Num produces an ionization
map which is visually more similar to that of \ctwo. On the other
hand, the \HII regions in the ionization maps produced by CPS+GS appears
to be more connected than the other two simulations. This is again a
indication of the fact that the ionization maps are more strongly
correlated with the density fields in case of CPS+GS than for the
other two simulations. We investigate this and other differences in
further details in the following sections using various statistical
measures.

\subsection{Bubble Size Distribution}
\label{sec:bubble-size} 
The bubble size distribution is one of the basic measures of the
morphology of the ionization fields.  However, due to the complex
three-dimensional morphology of the ionized regions, no unique method
exists that captures this distribution. Instead, several different
methods exist, which each show different properties of the size
distribution of ionized regions \citep{friedrich11}. Here we compare
the results of our three simulations using three of these, namely the
spherical average method, the friends-of-friends bubble size
distribution and the (spherically averaged) power spectrum of the
ionization field.

\begin{figure*}
\psfrag{z=10.290, xh1=0.898}[c][r][1][0]{{$\xb=0.90$}}
\psfrag{z=9.164, xh1=0.557}[c][r][1][0]{{$\xb=0.56$}}
\psfrag{z=8.892, xh1=0.375}[c][r][1][0]{{$\xb=0.38$}}
\psfrag{z=8.636, xh1=0.147}[c][r][1][0]{{$\xb=0.15$}}
\psfrag{C2RAY}[r][r][1][0]{{\ctwo}}
\psfrag{Semi-Num e=0.0}[r][r][1][0]{{Sem-Num}}
\psfrag{21cmFASTL}[r][r][1][0]{{CPS+GS}}
\psfrag{CPS-NG}[r][r][1][0]{{CPS}}
\psfrag{V dP/dV}[c][c][1][0]{{\small $V dP/dV$}}
\psfrag{V (Mpc3)}[c][c][1][0]{{\small $V\, ({\rm Mpc}^3)$}}
\includegraphics[width=1.\textwidth,
  angle=0]{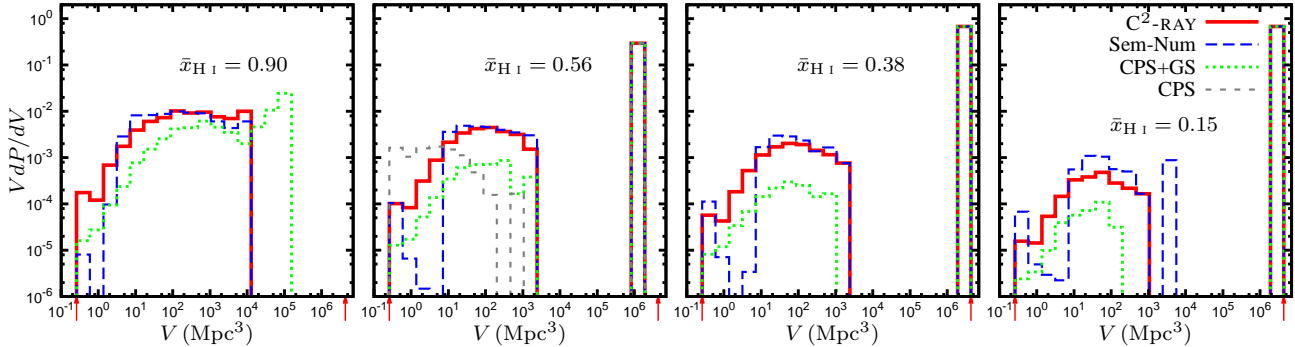}

  \caption{The FoF bubble size distribution at four representative
    values of $\xb$. The red arrows mark the cell and the box
    volume. We show the distribution for CPS at $\xb=0.56$ only, to
    illustrate its marked difference from \ctwo.}
\label{fig:fof_spec}
\end{figure*}
\subsubsection{Spherical average method}
First we use a spherical average technique, similar to \citet{zahn11}
and \citet{friedrich11}, to estimate the bubble size distribution from
different simulations. Figure \ref{fig:spa_spec} shows the bubble
radius distribution $(R dP/dR)$ for the three different simulations at
four representative stages of reionization. The distribution have some
characteristics which are common to all three simulations. During the
early stages of EoR all three simulations produce relatively smaller
bubbles, the maximum bubble size is restricted to $R \leq 10\,{\rm
  Mpc}$ and the peak of the distribution appears around $R \simeq
1\,{\rm Mpc}$. As reionization progresses the peak of the distribution
gradually shifts towards larger bubble size and finally reaches $R
\simeq 50\,{\rm Mpc}$ at the end stages of EoR.

Except at very early stages of EoR all three simulations show good
agreement in their spherical average bubble size distribution
throughout the history of reionization for most of the length
scales. There is almost no difference between the two semi-numerical
schemes ({\it i.e.} Sem-Num and CPS+GS). During very early stages of
EoR ($\xb = 0.90$), distribution for both semi-numerical simulations
differ slightly from \ctwo for larger bubble radii ($R \sim 2-10\,{\rm
  Mpc}$) and for very small bubble size (comparable to the grid
size). Except this during all other stages of reionization, \ctwo
differs from both semi-numerical simulations only at very small length
scales.

For comparison we also show the bubble size distribution for CPS
(without Gaussian smoothing) at the stage when $\xb = 0.56$.  We find
that the distribution in this case is significantly different even
from CPS+GS at both large and small scales ($R \sim 10\,{\rm
  Mpc}$). The $R dP/dR$ appears to be biased towards small scales,
whereas excess small scale bubble production is compensated by very
low bubble population at the large scales. This essentially indicates
a much stronger ``inside-out'' reionization than CPS+GS in this case.

\citet{zahn11} obtained similar results with this technique. They
found that, due to its over-connected nature, FFRT (equivalent of
CPS+GS in our case) produces more large scale bubbles and fewer small
scale bubbles compared with a radiative transfer simulation. The other
semi-numerical scheme \citep{mesinger07} in their analysis which is
similar to Sem-Num, also shows similar behaviour. This is probably due
to the fact that contrary to Sem-Num, which identifies only the
central pixel of the smoothing sphere as ionized when
eq. (\ref{eq:sem-num-ion}) is satified, \citet{mesinger07} identifies
the entire smoothing sphere to be ionized. This leads to the
production of more large scale ionized regions. Also, \citet{zahn11}
have tuned their semi-numerical schemes so as to achieve the same
evolution of the volume-averaged neutral fraction as that of the
radiative transfer simulations, whereas we have compared different
simulations at the level of same mass-averaged neutral fraction. This
in addition may enhance these small discrepancies. In addition to this
we would also like to note that previous studies by
\citet{friedrich11} have shown that the spherical average technique
tends to wash out some of the complicated features in the bubble shape
(most of which are essntially non-spherical) and produce smoother
distributions. We confirm this behaviour here as well.

\subsubsection{Friends-of-friends analysis}
We next use a friends-of-friends (FoF) algorithm, as in
\citet{iliev06a} and \citet{friedrich11}, to identify ionized regions
from our simulations. In this method, for a gridded ionization map,
two adjacent grid cells are considered to be part of the same ionized
region if they fulfill the same condition. Here, we use an ionization
threshold condition of $x_{th} \ge 0.5$. One important characteristic
of this method is that it does not presume anything about the shape of
the ionized regions and literally allows connected ionized regions of
any shape to be identified.

For a field consisting of continuous ionization fractions, such as
produced by a numerical simulation, the results of this method depend
on the choice for the ionization threshold, as shown in
\citet{friedrich11}. For the semi-numerical simulations, in which the
ionization fraction is either $0$ or $1$ the FoF is defined
uniquely. However, one may wonder if even in this case the results are
very sensitive to small scale features, either connecting or not
connecting two ionized regions. We tested this by performing the same
analysis after applying a Gaussian or a spherical smoothing filter
with the effective width of $3$ simulation cells to the ionization
fields. We found that this procedure, as expected, does reduce the
number of small bubbles. However, it does not significantly affect the
distribution at intermediate and large length scales. We therefore
conclude the FoF statistics for volumes above $100\,{\rm Mpc}^3$ to be
a robust result and not sensitive to small scale effects.

Figure \ref{fig:fof_spec} shows the bubble volume distribution ($V
dP/dV$) for the three simulations at four representative stages of
reionization. The distributions have some characteristics which are
common to all three simulations. One of the main features is that the
distributions are not continuous except at the very early stages of
reionization ($\xb \ge 0.90$). Another distinct feature is that once
the early phase of reionization is over there is one connected large
\HII region which is comparable to the volume of the simulation box
($\sim 10^6\,{\rm Mpc}^3$). The rest of the \HII regions are much
smaller ($\sim 0.26 - 10^4\, {\rm Mpc}^3$) in size and have an almost
continuous distribution for all simulations.

The bubble size distribution for Sem-Num is quite similar to that of
\ctwo during almost the entire period of reionization.  The only
disparity in the bubble size distribution between the two appears at
relatively small length scales. The number of bubbles is notably lower
in the Sem-Num simulation for the volume range $\sim 1-10 \,{\rm
  Mpc}^3$ than for \ctwo. For the smallest volume bin this number is
slightly larger than \ctwo.

In contrast, CPS+GS produces clearly quite different results. This is
especially evident at the intermediate and smaller length scales where
it produces significantly fewer bubbles than \ctwo and Sem-Num
do. This is consistent with our previous observation of the ratio
$\langle \xh1 \rangle_{v}/\langle \xh1 \rangle_{m}$ in Figure
\ref{fig:xh1_his}. We observe that at all stages of the EoR $\langle
\xh1 \rangle_{v}/\langle \xh1 \rangle_{m}$ is higher for CPS+GS than
for the other two simulation, {\it i.e.} the volume fraction of \HI is
always more than the mass fraction of \HI in CPS+GS. This implies that
reionization is more biased around high density regions in CPS+GS than
the other two schemes (more ``inside-out'').

For the purpose of comparison here also we show the bubble size
distribution for CPS (without Gaussian smoothing) at the stage when
$\xb = 0.56$. We find that the distribution in this case is markedly
different even from CPS+GS. The $VdP/dV$ is significantly biased
towards small scales, whereas excess small scale bubble production is
compensated by almost zero bubble population at the intermediate
scales. This indicates a much stronger ``inside-out'' reionization
than CPS+GS in this case.  At the same reionization state CPS+GS
generates considerably fewer bubbles at the smallest scales
(comparable to the cell volume). This is probably due to the fact that
the density field in case of the CPS+GS is more diffused which
prevents the over-production of very small ionized regions.

Furthermore, the way the two semi-numerical simulations treat
recombinations can also affect the bubble size distributions. In both
of the semi-numerical methods we assume a uniform recombination rate
throughout the IGM, which can be considered to be equivalent to having
no recombinations at all (as they can be effectively absorbed in the
source efficiency parameter $N_{\rm ion}$ or $\zeta$). This inaccurate
implementation of recombinations can lead to the discrepancy in bubble
size distribution. However, we will see in the following sections that
this does not drastically affect the simulated 21-cm signal from these
semi-numerical simulations. This is because the 21-cm signal is a
product of neutral fraction and density fluctuations. During the early
stages of the EoR it is the density fluctuations which plays a
dominating role over the fluctuations in $\xh1$, thus reduces the
effect of differences in the ionization maps at this stage.

\begin{figure*}
\psfrag{C2RAY}[r][r][1][0]{{\ctwo}}
\psfrag{Sem-Num e=0.0}[r][r][1][0]{{Sem-Num}}
\psfrag{21cmFASTL}[r][r][1][0]{{CPS+GS}}
\psfrag{CPS-NG}[r][r][1][0]{{CPS}}
\psfrag{z=10.290, xh1=0.898}[c][c][1][0]{{$\xb=0.90$}}
\psfrag{z=9.164, xh1=0.557}[c][c][1][0]{{$\xb=0.56$}}
\psfrag{z=8.892, xh1=0.375}[c][c][1][0]{{$\xb=0.38$}}
\psfrag{z=8.636, xh1=0.147}[c][c][1][0]{{$\xb=0.15$}}
\psfrag{k3 Pxx(k)}[c][c][1][0]{$k^3 P_{xx}(k)/(2\pi^2)$}
\psfrag{k (Mpc)}[c][c][1][0]{$k\, ({\rm Mpc}^{-1})$}
\includegraphics[width=1\textwidth,
  angle=0]{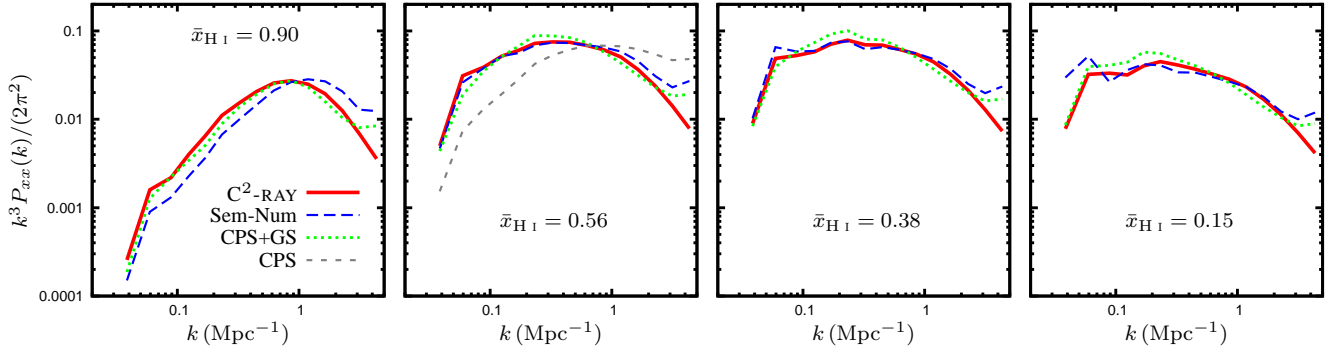}

  \caption{The power spectrum of the ionization field
    $P_{xx}(k)$ at four representative neutral fraction values. We
    show the power spectrum for CPS at $\xb=0.56$ only, to
    illustrate its marked difference from \ctwo.}
\label{fig:pk_ion_spec}
\end{figure*}
\subsubsection{Power spectrum}
\label{sec:PS}
The power spectrum of the ionization field ($P_{xx}(k)$) is also a
measure of the \HII bubble size distribution. It measures the
amplitude of fluctuations in the ionization field at different length
scales. It also directly contributes to the redshift space 21-cm \HI
power spectrum, which is a major observable quantity for the present
and the future EoR experiments. Figure \ref{fig:pk_ion_spec} shows the
power spectrum of the ionization maps at four representative stages of
the EoR. The power spectra of the ionization field also have some
features common to all three simulations of reionization considered
here. Similar to the bubble size distribution, the peak of the power
spectrum gradually shifts from small to large length scales ({\it
  i.e.} from large to small $k$ modes) and its amplitude also
increases as reionization progresses (up to $\xb \ge 0.5$). This
indicates the gradual growth and merger of the \HII regions with
time. Finally, at the very late stages of the EoR, when most of the
volume is ionized, the power spectrum becomes almost flat and there is
a significant decrement in its overall amplitude as well.

The differences in power spectrum between the three different
simulations is relatively small. At the very early stages of
reionization, Sem-Num produces less power at almost all scales (except
very small scales) compared with the other two simulations, whereas
CPS+GS is in good agreement with \ctwo at this stage.  At the
intermediate and late stages of the EoR, Sem-Num replicates the power
spectrum from \ctwo much better than CPS+GS, except at the very small
length scales. However at the smallest length scales, Sem-Num always
produces more power than the other two simulations. This is due to the
fact that it produces more small scale bubbles than the other two
schemes, which is also evident in the bubble size distributions
(Sect. \ref{sec:bubble-size}). Overall during the intermediate stages
of EoR, $P_{xx}(k)$ for Sem-Num lies within $15\%$ of that of \ctwo,
whereas for CPS+GS it lies within $25\%$ of that of \ctwo (for $k \leq
2.0\, {\rm Mpc}^{-1}$). As reported in previous studies
\citep{zahn11,friedrich11}, we also note that the power spectrum
analysis essentially produce similar results as that of the spherical
average technique.

For comparison we also show the power spectrum from CPS when
reionization is almost half way through ($\xb \approx 0.56$). We
observe that as we do not use a diffused density field in CPS, it
produces more small \HII regions than CPS+GS. This means more power at
small scales and less power at large scales, which changes the shape
of the power spectrum significantly. In a similar analysis,
\citet{zahn11} find that their FFRT scheme produces more power at both
largest and smallest scales than the radiative transfer schemes. This
we do not encounter in case of CPS+GS. Due to the significant
differences observed in the history, bubble size distribution and
power spectrum between CPS (without Gaussian smoothing) and other
simulations, we drop it from our comparison analysis here onwards.

\begin{figure*}
\psfrag{r}[c][c][1][0]{{${\mathcal R}_{xx}(k)$}}
\psfrag{k (Mpc)}[c][c][1][0]{{$k \,({\rm Mpc}^{-1})$}}
\psfrag{xh1=0.898}[c][c][1][0]{{$\xb=0.90$}}
\psfrag{xh1=0.557}[c][c][1][0]{{$\xb=0.56$}}
\psfrag{xh1=0.375}[c][c][1][0]{{$\xb=0.38$}}
\psfrag{xh1=0.147}[c][c][1][0]{{$\xb=0.15$}}
\psfrag{Sem-Num}[r][r][1][0]{{Sem-Num}}
\psfrag{CPS}[r][r][1][0]{{CPS+GS}}
\includegraphics[width=1.0\textwidth,
  angle=0]{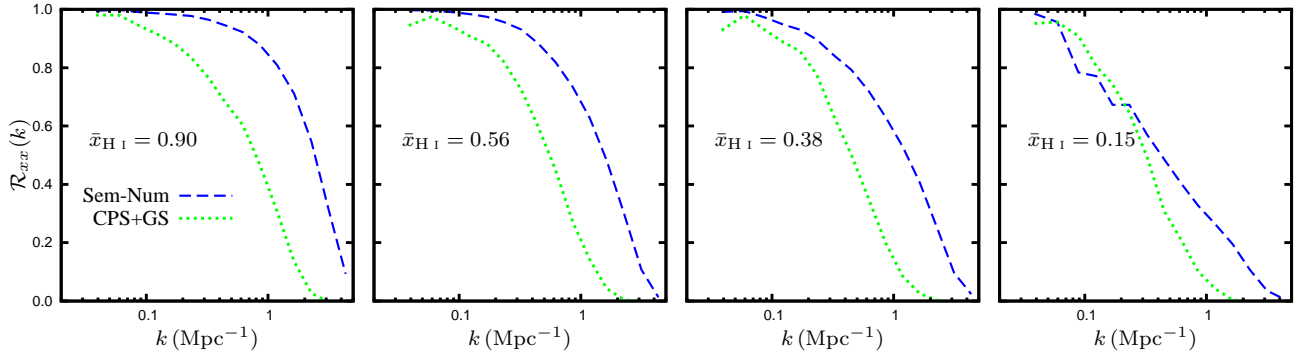}
  \caption{The cross-correlation ${\mathcal R}_{xx}(k)$ of
    the ionization maps of different semi-numerical schemes with the
    \ctwo simulation.}
\label{fig:comp_cross}
\end{figure*}
\begin{figure*}
\psfrag{C2RAY}[r][r][1][0]{{\ctwo}}
\psfrag{Sem-Num e=0.0}[r][r][1][0]{{Sem-Num}}
\psfrag{21cmFASTL}[r][r][1][0]{{CPS+GS}}
\psfrag{z=10.290, xh1=0.898}[c][c][1][0]{{$\xb=0.90$}}
\psfrag{z=9.164, xh1=0.557}[c][c][1][0]{{$\xb=0.56$}}
\psfrag{z=8.892, xh1=0.375}[c][c][1][0]{{$\xb=0.38$}}
\psfrag{z=8.636, xh1=0.147}[c][c][1][0]{{$\xb=0.15$}}
\psfrag{r}[c][c][1][0]{$r_{\Delta x}(k)$}
\psfrag{k (Mpc)}[c][c][1][0]{$k\, ({\rm Mpc}^{-1})$}
\includegraphics[width=1.\textwidth,
  angle=0]{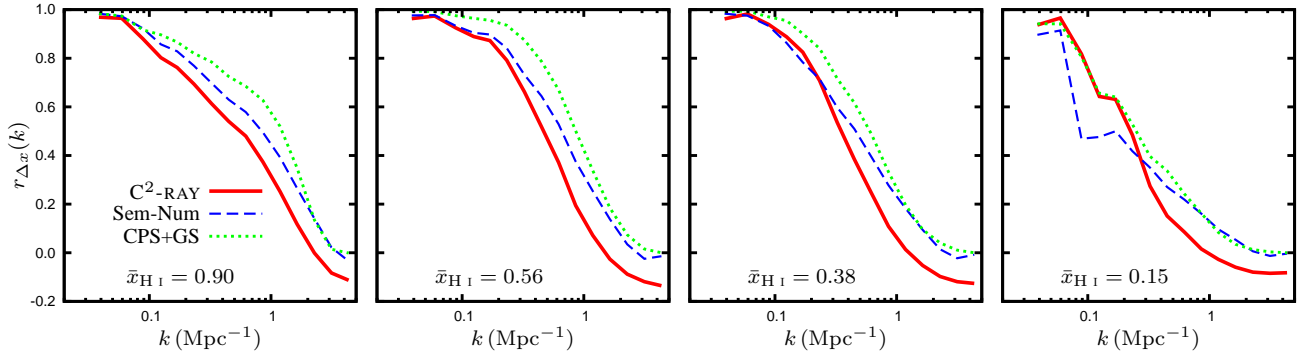}
  \caption{The cross correlation $r_{\Delta x}(k)$ between the
    ionization and the density field at four representative stages of
    reionization.}
\label{fig:r_spec}
\end{figure*}
\subsubsection{Comparison of bubble size distribution results} 
From these three methods for bubble size distributions we conclude
that the two methods which do not take into account connectivity,
namely the spherical average technique and the power spectrum
analysis, show that apart from small scales, there is good agreement
between all three simulation results, Sem-Num lying within 15\% and
CPS+GS within 25\% of the \ctwo results. The FoF method, which focuses
on connectivity, shows that the Sem-Num and \ctwo results agree quite
well, but CPS+GS shows fewer small and intermediate size \HII regions,
something which is also visible in Figure \ref{fig:his_map}. This is
partly due to the more inside-out nature of CPS+GS which means that a
given mean ionized mass fraction corresponds to a smaller ionized
volume fraction, but also because the ionized regions are more
connected and more quickly merge to form larger ionized regions.

\subsection{Cross-correlation}
\label{sec:cross-corr}
The cross-correlation between two different ionization fields A and B,
defined as ${\mathcal R}_{xx}(k) = P_{AB}(k)/\sqrt{P_{AA}(k)
  P_{BB}(k)}$, estimates how spatially correlated the two fields
are. We use ${\mathcal R}_{xx}(k)$ to quantify the strength of
correlation between a semi-numerical simulation and \ctwo at different
length scales. Figure \ref{fig:comp_cross} shows ${\mathcal
  R}_{xx}(k)$ estimated at four representative stages of EoR. From
this figure it is evident that ionization maps from both of these
semi-numerical schemes are highly correlated (${\mathcal R}_{xx} \ge
0.95$) with that of \ctwo at sufficiently large length scales ($k \leq
0.1 \,{\rm Mpc}^{-1}$), throughout almost all the stages of the
EoR. Also, one of the main common features of these two
cross-correlation coefficients is that they gradually decline at
smaller length scales with the progress of reionization. However, this
decline is faster for CPS+GS than for Sem-Num. In almost all stages of
the EoR the cross-correlation between Sem-Num and \ctwo is ${\mathcal
  R}_{xx} \ge 0.75$ at $k \le 1.0 \,{\rm Mpc}^{-1}$, whereas at the
same length scale range the cross-correlation between CPS+GS and \ctwo
can become as low as ${\mathcal R}_{xx} \sim 0.1$. This
cross-correlation analysis therefore further confirms the result that
the morphology of the ionization fields obtained from Sem-Num
resembles more that of \ctwo than CPS+GS. Our results are consistent
with the findings of \citet{zahn11} in this regard.

We also estimate the cross-correlation between the ionization and the
density field, defined as, $r_{\Delta x}(k) = P_{\Delta
  x}(k)/\sqrt{P_{xx}(k) P_{\Delta \Delta}(k)}$. This quantity tells us
how the distribution of ionized regions in different simulation
schemes are correlated with the underlying density field. Generally it
is expected that overdense regions in the density field will ionize
first as they are expected to host most of the ionizing sources. This
is known as ``inside-out'' reionization. The cross-correlation
coefficient $r_{\Delta x}(k)$ will directly quantify the strength of
this ``inside-out''-ness in different simulations. Figure
\ref{fig:r_spec} shows $r_{\Delta x}(k)$ for the three different
simulations of reionization that we have discussed here. One general
feature of $r_{\Delta x}(k)$ is that for all three simulations it is
highest at the largest length scales and gradually declines for
smaller scales. Also, the strength of $r_{\Delta x}$ is higher in the
early and the intermediate stages of EoR and smaller in the late
stages of reionization. Among the two semi-numerical schemes, at large
length scales ($k \le 0.7 \,{\rm Mpc}^{-1}$), Sem-Num follows \ctwo
more closely than CPS+GS at almost all stages of the EoR.

The cross-correlation coefficient $r_{\Delta x}(k)$ is always highest
for CPS+GS compared with the other two simulations, for all length
scales and in all stages of the EoR. This shows that the CPS+GS 
is the most inside-out in nature among all three
simulations discussed here. In other words, the ionization field of
CPS+GS traces the matter distribution more closely than \ctwo and
Sem-Num, which further supports our earlier observations. Similarly,
we observe that \ctwo is the least inside-out among the three
schemes. Sem-Num lies somewhere in between \ctwo and CPS+GS in terms
of its inside-out nature. The cross-correlation coefficient $r_{\Delta
  x}$ for Sem-Num follows that of \ctwo very closely up to the length
scales $k \le 0.7 \,{\rm Mpc}^{-1}$. For smaller length scales ({\it
  i.e.} higher $k$ values) it follows the CPS+GS. The strong
inside-out nature of the CPS+GS is in agreement with our earlier
observations of the evolution of its history, bubble size distribution
and the power spectrum of the ionization maps of this
simulation. These results are also consistent with the findings of
\citet{zahn11}.

%% file: signal.tex
\begin{figure*}
\psfrag{Mpc}[c][c][1][0]{\large{Mpc}}
\psfrag{mK}[c][c][1][0]{\large{$\delta T_b\,$ (mK)}}
\psfrag{C2RAY}[c][c][1][0]{\large{\ctwo}}
\psfrag{Sem-Num}[c][c][1][0]{\large{Sem-Num}}
\psfrag{CPS}[c][c][1][0]{\large{CPS+GS}}
\includegraphics[width=1.\textwidth,
  angle=0]{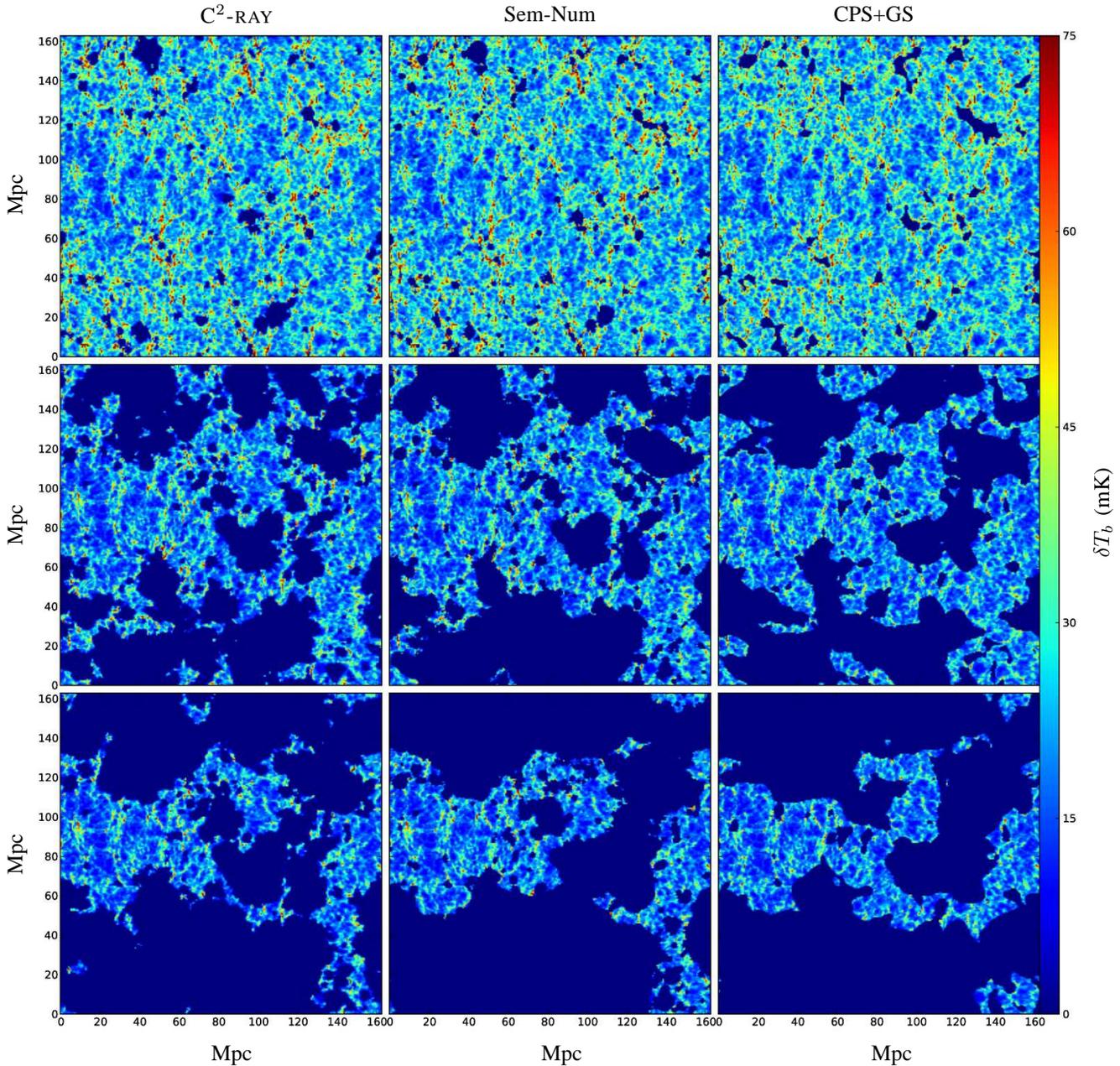}
  \caption{The redshift space 21-cm brightness temperature
    slices from three different simulations at three representative
    stages of reionization. The line of sight (LoS) is along the
    x-axis. From top to bottom three different rows correspond to $\xb
    = 0.90,\, 0.46$ and $0.26$ respectively.  Thickness of each slice
    is $0.64$ Mpc.}
\label{fig:dt_maps}
\end{figure*}
\begin{figure*}
\psfrag{r}[c][c][1][0]{{${\mathcal R}_{\delta T_b}(k)$}}
\psfrag{k (Mpc)}[c][c][1][0]{{$k \,({\rm Mpc}^{-1})$}}
\psfrag{xh1=0.898}[c][c][1][0]{{$\xb=0.90$}}
\psfrag{xh1=0.557}[c][c][1][0]{{$\xb=0.56$}}
\psfrag{xh1=0.375}[c][c][1][0]{{$\xb=0.38$}}
\psfrag{xh1=0.147}[c][c][1][0]{{$\xb=0.15$}}
\psfrag{Sem-Num}[r][r][1][0]{{Sem-Num}}
\psfrag{CPS}[r][r][1][0]{{CPS+GS}}
\includegraphics[width=1.0\textwidth,
  angle=0]{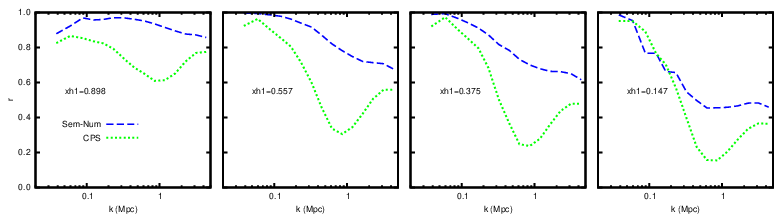}
  \caption{The cross-correlation ${\mathcal R}_{\delta T_b}
    (k)$ of the redshift space brightness temperature maps of two
    semi-numerical simulations with that of the \ctwo.}
\label{fig:comp_cross_rsd}
\end{figure*}

\section{Redshifted 21-cm Signal}
\label{sec:signal}
The major aim of the present and upcoming radio interferometric
surveys of the EoR is to detect the redshifted \HI 21-cm signal from
this epoch. Once detected, reionization simulations can be used to
interpret these observations. Hence it is very important to test
different semi-numerical schemes for their ability to simulate this
signal. The brightness temperature of the 21-cm \HI radiation from EoR
(when it can be assumed that the spin temperature is much higher than
the CMBR temperature, $T_S \gg T_{\gamma}$) can be expressed as
\begin{equation}
  \delta T_b ({\bf x}, z) =
  \overline{\delta T_b }(z) \left[ 1 + \delta_{\rho_{\HI}}({\bf
      x})\right]\,,
\label{eq:dT}
\end{equation}
where $\delta_{\rho_{\HI}}({\bf x})$ is the fluctuation in \HI density
at the point ${\bf x}$ and $\overline{\delta T_b }(z)$ is the mean
brightness temperature at redshift $z$. We estimate the brightness
temperature in real space from the ionization and the density fields
of our simulations following eq. (\ref{eq:dT}). 

\subsection{Redshift space distortions}
\label{sec:rsd}
Coherent inflows of matter (and gas) into overdense regions and
outflows of matter (and gas) from underdense regions make the observed
21-cm signal anisotropic along the line of sight (LoS). This
particular anisotropy in the signal is known as the redshift space
distortions. We next explain how we have implemented the effects of
the redshift space distortions on the brightness temperature maps
generated from the simulations. One of the most accurate methods to
include the effect of redshift space distortions is to include it at
the level of each individual simulation particle. In this method one
assumes that each particle from the $N$-body simulation contains an
equal amount of hydrogen mass ($M_H$) before any reionization has
actually taken place. The ionization map at each snapshot redshift
provides us with a neutral fraction $\xh1$ at each grid point of the
simulation box. For each individual simulation particle one can then
interpolate the neutral fraction from its eight nearest neighbouring
grid points to determine the neutral fraction at that particle's
position. This can be used to calculate the \HI mass of the $i$th
particle as $M^i_{\HI} = \xh1^i M_H$. Next, one considers a distant
observer located along the $x$ axis and the $x$ component of the
peculiar velocity $(v_x)$ of that particle is then used to determine
its location in redshift space as
\begin{equation}
s = x + \frac{v_x}{a H(a)} \,,
\label{eq:rsd}
\end{equation}
where $a$ and $H(a)$ are the scale factor and the Hubble parameter
respectively. Finally, one interpolates the \HI distribution from the
particles to the grid and uses that to estimate the 21-cm signal in
redshift space. This method of mapping the 21-cm signal from
real to redshift space is similar to the PPM-RRM method of
\citet{mao12} and also the method described by \citet{majumdar13}.

However, this method becomes computationally very expensive when one
has to deal with a large number of particles ({\it e.g.} $3072^3$
particles in our case). Therefore, instead of this particle based
method we use the grid based method discussed in \citet{jensen13} to
avoid this problem. In this method we divide each grid cell into $n$
equal sub-cells along the LoS. If the brightness temperature of the
original grid cell was $\delta T_b ({\bf x})$, then each sub-cell is
assigned with a brightness temperature $\delta T_b ({\bf x})/n$. We
then interpolate the velocity and density fields onto the sub-cells
and move them according to eq. (\ref{eq:rsd}) and map the redshift
space $\delta T_b$ to the original grid again. For all the simulations
described in this paper, we have used $50$ sub-cells along the LoS for
each original grid cell (of size $0.64$ Mpc) to implement the redshift
space distortions. This technique is somewhat similar to the MM-RRM
method described in \citet{mao12}. Figure \ref{fig:dt_maps} shows the
brightness temperature maps in redshift space for the three different
simulations discussed here at three different stages of the EoR.

\subsection{Cross-correlation}
\label{sec:sig-cross}
Redshift space distortions will change the 21-cm signal along the LoS.
It is thus important to compare the simulations discussed here in
their ability to predict the redshifted 21-cm brightness temperature
fluctuations as well as various other statistical measures of it in
redshift space. We estimate the cross-correlation coefficient
${\mathcal R}_{\delta T_b}(k)$ between the redshift-space brightness
temperature maps of the two semi-numerical simulations with that of
\ctwo, to quantify how well the signal is reproduced by these
semi-numerical schemes at different length scales. Figure
\ref{fig:comp_cross_rsd} shows this cross-correlation coefficient
${\mathcal R}_{\delta T_b}(k)$ at different stages of the EoR. The
overall characteristics of ${\mathcal R}_{\delta T_b}(k)$ shows that
the semi-numerical schemes are more strongly correlated with \ctwo at
the early stages of EoR than at the late stages and the correlation is
higher at larger length scales and gradually declines towards smaller
length scales.

We also observe that ${\mathcal R}_{\delta T_b}(k)$ for both of the
semi-numerical schemes is much higher at all scales compared with
${\mathcal R}_{xx}(k)$ (the cross-correlation between ionization
fields; see Figure \ref{fig:comp_cross}). A possible reason for this
is the following: The brightness temperature fluctuations $\delta T_b$
are a combination of fluctuations in both the density field and the
neutral fraction (see eq. [\ref{eq:dT}]). All three simulations have
the same density fluctuations (note that CPS+GS has a slightly
diffused density field compared to the others) and they differ only in
their ionization maps.  In addition, for a completely neutral medium,
redshift space distortions will effectively add some fluctuations to
$\delta T_b$ that are related to the density fluctuations. At the
early stages of reionization the fluctuations in $\delta T_b$ maps
will thus be strongly dominated by density fluctuations rather than by
fluctuations in the ionization field. This will make ${\mathcal
  R}_{\delta T_b}(k)$ higher than than ${\mathcal R}_{xx}(k)$. As
reionization progresses, this dominance of density fluctuations will
be gradually taken over by the fluctuations in the ionization
maps. This will effectively reduce the cross-correlation ${\mathcal
  R}_{\delta T_b}$ at all scales.

Among the two semi-numerical simulations, Sem-Num provides a better
reproduction of the signal than CPS+GS at almost all length scales and
in all stages of reionization. For Sem-Num, ${\mathcal R}_{\delta T_b}
\ge 0.7$ for $k \le 1\, {\rm Mpc}^{-1}$ in almost all stages of
reionization, whereas the same for CPS+GS is ${\mathcal R}_{\delta
  T_b} \ge 0.25$. At larger length scales ($k \ge 0.1\, {\rm
  Mpc}^{-1}$) the correlation is even stronger (${\mathcal R}_{\delta
  T_b} \ge 0.95$) for Sem-Num whereas for CPS+GS, ${\mathcal
  R}_{\delta T_b} \ge 0.8$. The value of ${\mathcal R}_{\delta T_b}$
for Sem-Num never goes below $0.6$ even at the smallest length scales
until the end stages of the EoR ($\xb \le 0.15$).  This
cross-correlation with \ctwo establishes the fact that among the two
semi-numerical schemes, Sem-Num provides a better representation of
the observable signal than CPS+GS.

\begin{figure*}
\psfrag{C2RAY}[r][r][1][0]{{\ctwo}}
\psfrag{Sem-num e=0.0}[r][r][1][0]{{Sem-Num}}
\psfrag{21cmFASTL}[r][r][1][0]{{CPS+GS}}
\psfrag{RSD}[r][r][1][0]{{Redshift Space}}
\psfrag{real}[r][r][1][0]{{Real Space}}
\psfrag{Variance of dT (mK2)}[c][c][1][0]{{$\sigma^2\,\,({\rm mK}^2)$}}
\psfrag{z}[c][c][1][0]{{$z$}}
\psfrag{xh1}[c][c][1][0]{{$\xb$}}
\includegraphics[width=.8\textwidth,
  angle=0]{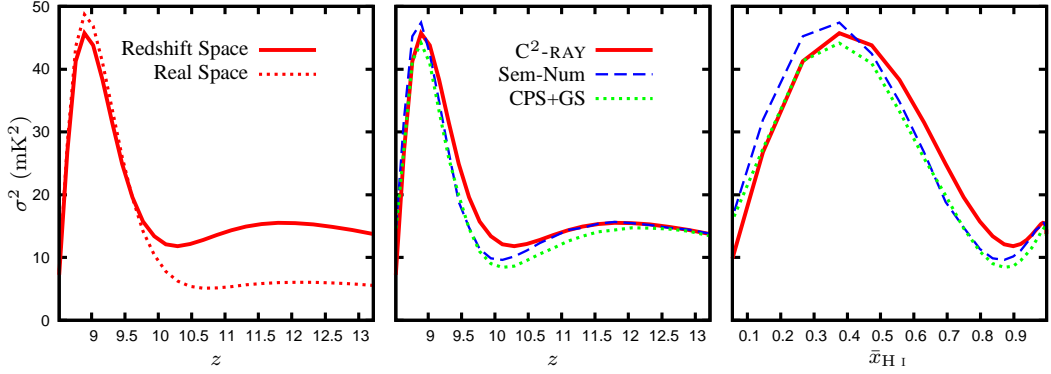}

  \caption{The evolution of the variance of the 21-cm brightness
    temperature. The left panel shows the evolution in real and
    redshift space with redshift as obtained from \ctwo. The central
    panel shows the evolution with $z$ in redshift space for all three
    simulations. The right panel shows the same evolution with the
    neutral fraction $\xb$. In all panels the variance has been
    estimated for a LOFAR-like baseline distribution and with a
    Gaussian approximation for the point spread function.}
\label{fig:tb_var}
\end{figure*}
\subsection{Comparison of Observable Quantities}
\label{sec:observables}
The cross-correlation ${\mathcal R}_{\delta T_b}(k)$, presented in the
previous section, shows that the semi-numerical schemes can provide a
very good estimation for the redshifted 21-cm signal even up to
considerably small length scales. However, neither the present ({\it
  e.g.} LOFAR, GMRT, 21CMA etc) nor the upcoming ({\it e.g.} SKA)
radio interferometric arrays are able to image the signal with a
precision comparable to the resolution of these simulations. LOFAR may
be able to image the IGM during the EoR at relatively large length
scales ($\ge 0.5^{\circ}$) \citep{zaroubi12} or the individual \HII
regions around very bright sources \citep{geil08,majumdar12,datta12},
but the focus of the first generation of 21-cm EoR experiments will be
on the statistical detection of the signal. In the following sections,
we compare how well our simulations can predict various statistical
measures of the 21-cm signal.

\subsubsection{Variance of the 21-cm brightness temperature
  fluctuations}
\label{sec:variance}
The variance of the 21-cm brightness temperature fluctuations
holds the promise to be the statistical quantity through which the
first detection of the EoR 21-cm signal may be possible. The
variance essentially measures the power spectrum of the signal
integrated over all observable wave numbers (or $k$ modes). Once
detected, in principle the variance can be parametrized to constrain
the reionization redshift and its duration. It is anticipated that
this might be achievable even with just $600$ hr of observation using
LOFAR \citep{patil14}.

Even for a very crude statistical measure of the EoR 21-cm signal,
like the variance, the accurate implementation of the effect of
peculiar velocities is important. We illustrate the effects of
redshift space distortions on this observable quantity in the left
panel of Figure \ref{fig:tb_var}. This figure compares the evolution
of the variance of the signal in real and redshift space, simulated
using \ctwo. The variance shown here has been calculated for a
LOFAR-like baseline distribution and with a Gaussian approximation for
the point spread function of size $\sim 3.25'$ at $150$ MHz in slices
of thickness $38$ kHz in frequency. Each volume contains $256$ of such
slices and we calculate the final variance of $\delta T_b$ as the
average over the variance of each of these slices, to reduce the
uncertainties due to sampling errors.

We observe that the redshift space distortions change both the shape
and the amplitude of the signal considerably during the early stages
of reionization ({\it i.e.} for $z \ge 9.8$ and $\xb \ge 0.8$ in this
case). The amplitude of the variance in redshift space becomes
significantly higher at this stage ($\ge 2.5$ times more with respect
to the real space signal at $z \ge 10.6$ and $\xb \ge 0.94$). The
redshift space signal also shows a broad peak at $z \sim 11$ and $\xb
\sim 0.97$, whereas no such peak is visible in the real space
signal. The redshift space variance has a very prominent dip at $z
\sim 10.3$ and $\xb \sim 0.9$, which is not visible in its real space
counter part. This sharp decrement of the signal in the redshift space
is probably a signature of the negative contribution from the
cross-correlation between the density and the ionization
field. According to the linear \citep{barkana05} as well as the
quasi-linear \citep{mao12} models of the redshift space distortions,
this cross-correlation contributes more in the redshift space than in
real space. All of these together increases the possibility of
detection of the redshift space signal through the estimation of its
variance. This broad peak and the sharp dip in the variance of the
redshift space signal during the early stages of EoR has been ignored
in the variance analysis of \citet{patil14}.  It is also worth
mentioning that when the effect of peculiar velocities are
incorporated in a perturbative fashion similar to \citet{santos10} and
\citet{mesinger11} (as well as in \citealt{patil14}), it introduces an
additional error of $\ge 20\%$ in the signal \citep{mao12}. Thus it is
important to take into account the effect of the peculiar velocities
accurately when parametrization of the observed variance is done on
the basis of simulations. In the later stages of the EoR ({\it i.e.}
$\xb \le 0.8$), the redshift space variance does not show any
significant difference with its real space counterpart (deviation is
$\le 5\%$).

Next, we compare the predicted variance in redshift space from the
three different simulations considered in this work. The central and
the right panel of Figure \ref{fig:tb_var} show this comparison
through the evolution of the variance with redshift and $\xb$,
respectively. We observe that both of the semi-numerical simulations
follow \ctwo very well. During the early phase of EoR ({\it i.e.} $z
\ge 10.5$ and $\xb \ge 0.92$), the Sem-Num follows the \ctwo more
closely (deviation $\le 8\%$) than the CPS+GS (deviation $\le
20\%$). The deviation of the semi-numerical simulations from that of
\ctwo is more prominent during the intermediate stages of
reionization. This deviation from \ctwo peaks ($\sim 30\%$ for CPS+GS
and $\sim 20\%$ for Sem-Num) near the point (around $z \sim 10.1$ and
$\xb \sim 0.87$) where the variance shows a sharp dip. We have
found that among all three simulations CPS+GS is the most
inside-out ($r_{\Delta x}$ in Figure \ref{fig:r_spec}) in nature. This
implies that at this point the contribution of $r_{\Delta x}$ will be
largest for CPS+GS, which will result in a much sharper dip in
the variance predicted by this scheme. However, in the later stages of
the EoR ({\it i.e.} $z \le 9.6$ and $\xb \le 0.75$) the variance
predicted by both of the semi-numerical simulations stay within $\sim
10\%$ of that of \ctwo. Thus it is the differences in the source
models among the different semi-numerical schemes which causes the
differences in the variance predicted by them.
\begin{figure*}
\psfrag{C2RAY}[r][r][1][0]{{\ctwo}} 
\psfrag{Sem-Num e=0.0}[r][r][1][0]{{Sem-Num}}
\psfrag{21cmFASTL}[r][r][1][0]{{CPS+GS}} 
\psfrag{z=10.290, xh1=0.898}[c][c][1][0]{{$\xb=0.90$}} 
\psfrag{z=9.164, xh1=0.557}[c][c][1][0]{{$\xb=0.56$}} 
\psfrag{z=8.892, xh1=0.375}[c][c][1][0]{{$\xb=0.38$}} 
\psfrag{z=8.636, xh1=0.147}[c][c][1][0]{{$\xb=0.15$}} 
\psfrag{k3 Ph1(k)}[c][c][1][0]{$k^3 P^s_0(k)/(2\pi^2)\,({\rm mK^2})$}
\psfrag{k (Mpc)}[c][c][1][0]{$k\, ({\rm Mpc}^{-1})$}

\includegraphics[width=1\textwidth,
angle=0]{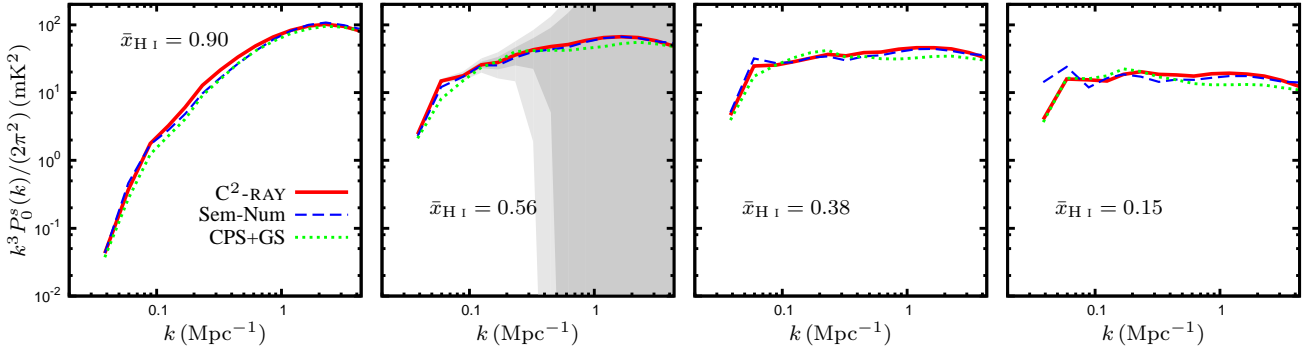}

\caption{The spherically averaged power spectrum of the redshift space
  21-cm signal. The shaded regions in light and dark gray represent the
  uncertainty due LOFAR-like system noise at $150$ MHz for $1000$ and
  $2000$ hr of observation respectively.}
\label{fig:pk_h1_spec}
\end{figure*}

\subsubsection{The redshift space 21-cm power spectrum and its angular
  multipole moments} 
\label{sec:rsd-PS}
Redshift space distortions make the 21-cm signal from
the EoR anisotropic. Thus the
power spectrum will depend on both the wave number $k$ and the
quantity $\mu={\bf k} \cdot {\bf \hat{n}}/k$, which is the cosine of
the angle between the wave vector ${\bf k}$ and the unit vector ${\bf
  \hat{n}}$ along the LoS \citep{kaiser87}. It is convenient
to decompose the anisotropy using Legendre polynomials $ {\mathcal
  P}_l(\mu)$ \citep{hamilton92,cole95} as
\begin{equation}
P^s(k,\mu) = \sum_{l \,{\rm even}} {\mathcal P}_l(\mu)
P_{l}^s(k)\, ,
\label{eq:p_sl_sum}
\end{equation}
where $P_{l}^s(k)$ are the different angular multipoles of
$P^s(k,\mu)$. This decomposition of the anisotropy is a representation
in an orthonormal basis. Thus different angular multipole moments in
this representation are orthogonal to each other
\citep{majumdar13}. The anisotropic power spectrum can also be
decomposed in different coefficients of the powers of $\mu$
\citep{mao12}. However these coefficients will not be completely
independent of each other and the correlation between them (or the
leakage of power from one component to the other) sometimes may give
rise to slightly wrong interpretations \citep{jensen13}. On the basis
of these angular multipole moments of the redshift space power
spectrum we compare our semi-numerical schemes with \ctwo.  As far as
we know, such a comparison has not been made before.

We Fourier transform the brightness temperature data cubes obtained
from different simulations and estimate the angular multipoles $P^s_l$
of the redshifted 21-cm power spectrum from the Fourier transformed
data following the equation
\begin{equation}
  P^s_{l} (k) = \frac{\left( 2l + 1 \right)}{4 \pi} \int {\mathcal P}_l(\mu)\, P^s (k)\, d\Omega \,,
\label{eq:p_sl}
\end{equation}
where $P^s (k)$ is the 21-cm power spectrum in redshift space. The
integral is done over the entire solid angle to take into account all
possible orientations of the ${\bf k}$ vector with the LoS direction
${\bf \hat{n}}$. Each angular multipole is estimated at $15$
logarithmically spaced $k$ bins in the range $0.038\leq k \leq 4.90\,
{\rm Mpc}^{-1}$. It is obvious from eq. (\ref{eq:p_sl_sum}) and
(\ref{eq:p_sl}) that the $0^{\rm th}$ angular moment or the monopole
($P^s_0$) will be essentially the spherically averaged 3D power
spectrum in redshift space.

To better understand and interpret our results, we have considered two
models for the redshift space power spectrum. The first of these
uses the linear approximations described in \cite{barkana05} to 
express the redshift space power spectrum as:
\begin{align}
  P^s (k,\mu) = \overline{\delta T_b }^2(z) &\left[P_{\xh1 \xh1}
    (k) + 2 (1 + \mu^2) P_{\Delta \xh1} (k)\right.\nonumber\\
  &\left. + (1 + \mu^2)^2 P_{\Delta \Delta} (k) \right]
\label{eq:lmodel}
\end{align}
where $\Delta_{\xh1}$ and $\Delta$ are the Fourier transform of the
neutral fraction and the density fluctuations and $P_{\xh1\xh1}$,
$P_{\Delta \Delta}$ are the power spectra of these two quantities
respectively, and $P_{\Delta \xh1}$ is the cross power spectrum
between $\Delta$ and $\Delta_{\xh1}$. In this model only the first
three even angular moments of the redshift space power spectrum have
non-zero values
\begin{equation}
 P^s_0  = \overline{\delta T_b }^2(z) \left( \frac{28}{15} P_{\Delta \Delta} + \frac{8}{3} P_{\Delta \xh1} + P_{\xh1 \xh1} \right)  \,,
\label{eq:lmodel0}
\end{equation}
\begin{equation}
  P^s_2 = \overline{\delta T_b }^2(z) \left( \frac{40}{21} P_{\Delta \Delta} + \frac{4}{3} P_{\Delta \xh1} \right) \,,
\label{eq:lmodel2}
\end{equation}
\begin{equation}
P^s_{4} = \overline{\delta T_b }^2(z) \left( \frac{8}{35} \right) P_{\Delta \Delta}\,.
\label{eq:lmodel4}
\end{equation}
In the quasi-linear model of \citet{mao12}, the same redshift space
power spectrum can be expressed as
\begin{align}
  P^s (k,\mu) = \overline{\delta T_b }^2(z) &\left[P_{\rho_{\HI}
      \rho_{\HI}}(k) + 2 \mu^2 P_{\rho_{\HI} \rho_H} (k)
  \right.\nonumber\\ &\left. + \mu^4 P_{\rho_H \rho_H} (k) \right]
\label{eq:qlmodel}
\end{align}
where $\Delta_{\rho_{\HI}}$ and $\Delta_{\rho_H}$ are the Fourier
transform of the neutral and the total hydrogen density fluctuations
and $P_{\rho_{\HI} \rho_{\HI}}$, $P_{\rho_H \rho_H}$ are the power
spectra of $\Delta_{\rho_{\HI}}$ and $\Delta_{\rho_H}$ respectively,
and $P_{\rho_{\HI} \rho_H}$ is the cross power spectrum between
$\Delta_{\rho_{\HI}}$ and $\Delta_{\rho_H}$. Also in this case only
the first three even angular multipole moments will have non-zero
values
\begin{equation}
  P^s_0  = \overline{\delta T_b }^2(z) \left( \frac{1}{5} P_{\rho_H \rho_H} + \frac{2}{3} P_{\rho_{\HI} \rho_H} + P_{\rho_{\HI} \rho_{\HI}} \right)  \,,
\label{eq:qlmodel0}
\end{equation}
\begin{equation}
  P^s_2 = \overline{\delta T_b }^2(z) \left( \frac{4}{7} P_{\rho_H \rho_H} + \frac{4}{3} P_{\rho_{\HI} \rho_H} \right) \,,
\label{eq:qlmodel2}
\end{equation}
\begin{equation}
  P^s_{4} = \overline{\delta T_b }^2(z) \left( \frac{8}{35} \right) P_{\rho_H \rho_H}\,.
\label{eq:qlmodel4}
\end{equation}
All the simulations discussed here work with the underlying assumption
that the baryons follow the dark matter distribution. This essentially
implies that the density fluctuations $\Delta$ and the total hydrogen
(ionized + neutral) density fluctuations $\Delta_{\rho_H}$ are
essentially the same quantity. Thus their power spectra are also the
same ({\it i.e.} $P_{\Delta \Delta} = P_{\rho_H \rho_H}$). This means
that according to both of these models the $4^{\rm th}$ moment (or the
hexadecapole $P^s_4$) measures the density power spectrum.
\begin{figure*}
\psfrag{P2/P0}[c][c][1][0]{$P^s_2(k)/P^s_0(k)$}
\psfrag{z=10.290, xh1=0.898}[c][c][1][0]{{$\xb=0.90$}}
\psfrag{z=9.164, xh1=0.557}[c][c][1][0]{{$\xb=0.56$}}
\psfrag{z=8.892, xh1=0.375}[c][c][1][0]{{$\xb=0.38$}}
\psfrag{z=8.636, xh1=0.147}[c][c][1][0]{{$\xb=0.15$}}
\psfrag{C2RAY}[r][r][1][0]{{\ctwo}}
\psfrag{Sem-Num e=0.0}[r][r][1][0]{{Sem-Num}}
\psfrag{21cmFASTL}[r][r][1][0]{{CPS+GS}}
\psfrag{k (Mpc)}[c][c][1][0]{$k\, ({\rm Mpc}^{-1})$}

\includegraphics[width=1.0\textwidth,
  angle=0]{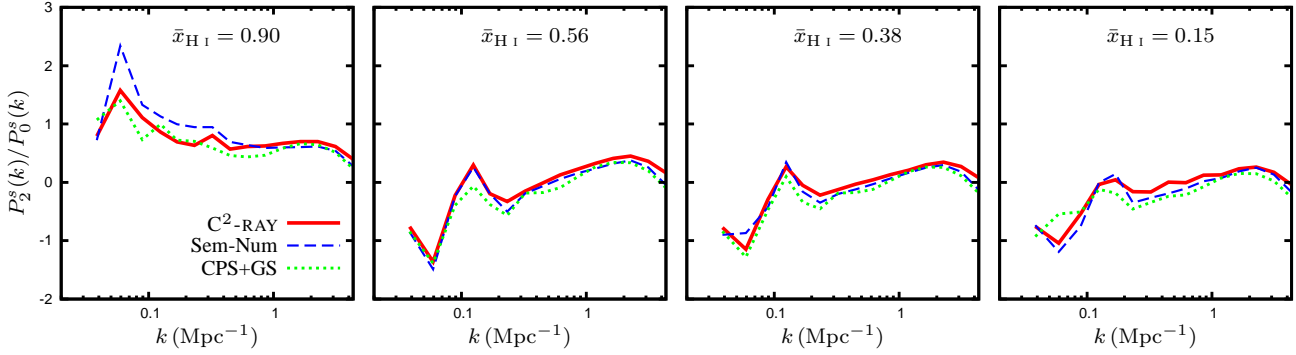}

  \caption{The ratio $P^s_2(k)/P^s_0(k)$ at four representative stages
    of the EoR.}
\label{fig:P2P0}
\end{figure*}
\begin{figure*}
\psfrag{P2/P0}[c][c][1][0]{\large{$P^s_2/P^s_0$}}
\psfrag{xh1}[c][c][1][0]{\large{$\xb$}}
\psfrag{k = 0.23 Mpc-1}[c][c][1][0]{$k = 0.23 \,{\rm Mpc}^{-1}$}
\psfrag{k = 0.059 Mpc-1}[c][c][1][0]{$k = 0.06 \,{\rm Mpc}^{-1}$}
\psfrag{k = 0.12 Mpc-1}[c][c][1][0]{$k = 0.12 \,{\rm Mpc}^{-1}$}
\psfrag{C2RAY}[r][r][1][0]{{\ctwo}}
\psfrag{Sem-Num e=0.0}[r][r][1][0]{{Sem-Num}}
\psfrag{21cmFASTL}[r][r][1][0]{{CPS+GS}}
\psfrag{toy-io}[r][r][1][0]{{Toy-inside-out}}
\psfrag{toy-oi}[r][r][1][0]{{Toy-outside-in}}
\includegraphics[width=1.0\textwidth,
  angle=0]{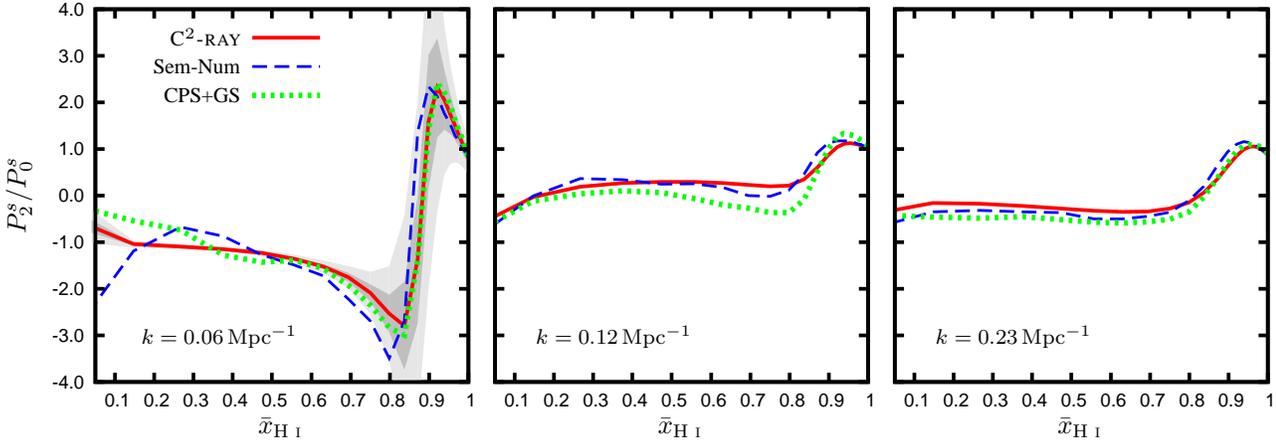}

  \caption{The evolution of the ratio $P^s_2/P^s_0$ with $\xb$ at
    three representative $k$ values. The shaded regions in light and
    dark gray represent uncertainty due to the system noise for $2000$ and
    $5000$ hr of observation using a LOFAR like instrument at $150$
    MHz.}
\label{fig:P2P0_speck}
\end{figure*} 

We first focus on the monopole moment ($P^s_{0}$) of the 21-cm
redshift space power spectrum ({\it i.e.} the spherically averaged
power spectrum), which measures the strength of the signal at
different length scales averaged over all angles. Figure
\ref{fig:pk_h1_spec} shows the monopole moment of the power spectrum
for different simulations at four representative stages of EoR. We
observe that the differences between the predictions for $P^s_{0}$
from \ctwo and the semi-numerical schemes is relatively small for
almost all stages of the EoR. In case of Sem-Num the predictions
deviates from \ctwo by $\le 10\%$ for most of the length scales. This
difference goes up to $20\%$ at most in some stages. In case of the
CPS+GS the $P^s_{0}$ deviates from that of the \ctwo by $\le 20\%$ for
most length scales at almost all stages and the difference can reach
$40\%$ at certain stages of the EoR. Overall, CPS+GS tends to
underestimate the power in some large and most of the small length
scales. This is probably a signature of the stronger correlation
between the density and the ionization field (see Figure
\ref{fig:r_spec}) in case of CPS+GS. According to both models of
redshift space distortions discussed above (eq. [\ref{eq:lmodel0}] and
[\ref{eq:qlmodel0}]) the cross-correlation power spectrum between the
density and the neutral fraction or the neutral density field
contributes negatively to $P^s_{0}$. Since the ionization map produced
using the CPS+GS is more strongly correlated with density field than
the other two simulations, this makes the amplitude of the $P^s_{0}$
lower in case of CPS+GS.

The shaded regions in the Figure \ref{fig:pk_h1_spec} (in the 2nd
panel from left) show the uncertainties in the measurements of the
$0^{\rm th}$ moment due to the system noise for a LOFAR like
instrument at $150$ MHz \citep{mcquinn06,datta12,jensen13}. It is
evident from this figure that even after $1000$ or $2000$ hr of
observation the signal will still possibly be dominated by the noise
for $k$ modes $\ge 0.35\, {\rm Mpc^{-1}}$ or $\ge 0.50\, {\rm
  Mpc^{-1}}$, respectively. Thus it would be of interest to see how
well the semi-numerical methods perform in predicting the signal for
$k$ values smaller than these limits. At these large length scales,
the $P^s_{0}$ estimated from both of the semi-numerical simulations
show significantly less difference from the same predicted by \ctwo.
This difference from \ctwo is less than $25\%$ for the CPS+GS and less
than $15\%$ for the Sem-Num for $k$ values below these upper limits.

The $2^{\rm nd}$ moment $P^s_2$ ({\it i.e.} the quadrupole moment)
essentially measures the anisotropy due to the peculiar velocities in
the signal. The presence of a measureable non-zero quadrupole moment
itself will be evidence of redshift space distortions. The ratio
between the quadrupole and the monopole moments [$P^s_2(k)/P^s_0(k)$]
of the 21-cm power spectrum can be used to quantify the strength as
well as the nature of the redshift space distortions present in the
observed signal. In principle it is possible to constrain the
reionization model if the nature and the evolution of this ratio
during the EoR can be probed with sufficient accuracy
\citep{majumdar13}. Figure \ref{fig:P2P0} shows the
$P^s_2(k)/P^s_0(k)$ estimated from the three different simulations at
four representative stages of EoR. One can easily figure out the
prominent general features of this observable quantity from this
figure. At the early stages of reionzation ($\xb \ge 0.9$) this ratio
is positive at all length scales. Once this phase is over ($\xb <
0.9$), it becomes negative at larger length scales ($\le 0.50\, {\rm
  Mpc^{-1}}$), due to the strong contribution from the
cross-correlation term (see eq. [\ref{eq:lmodel2}] and
[\ref{eq:qlmodel2}]). This ratio also developes a positive slope at
this stage of EoR, which gradually reduces as reionization progresses.
It is clear from Figure \ref{fig:P2P0} that most of these important
general features observed in \ctwo simulation are reproduced well by
both of the semi-numerical simulations discussed here. We further
compare the semi-numerical simulations with \ctwo in the length scale
range ($k \le 0.50\, {\rm Mpc^{-1}}$), where it is possible to detect
$P^s_0$. At the early stages of EoR for this length scale range,
CPS+GS produces a better match with that of the \ctwo (deviation is
$\le 15\%$), than Sem-Num (deviation is $\le 25\%$ and at $k \le
0.05\, {\rm Mpc^{-1}}$ it may go above $50\%$). However, after the
initial stages of EoR ($\xb \le 0.9$), the predictions from Sem-Num
(deviation is $\le 20\%$) are better matched with \ctwo, than CPS+GS
(deviation is $\le 30\%$). The major reason for this difference in
case of CPS+GS is possibly that the cross-correlation between the
density and the ionization field is much stronger in this case than in
Sem-Num or \ctwo.

To study this ratio in further detail, we show its evolution in Figure
\ref{fig:P2P0_speck} at three representative length scales ($k =
0.06,\,0.12\,{\rm and}\, 0.23\, {\rm Mpc^{-1}}$). For all three
simulations the evolution of this ratio can be characterised in
general by a sharp positive peak and a negative dip at the early phase
of EoR ($\xb \sim 0.9$). Once this early phase is over, $P^s_2/P^s_0$
remains negative for the remaining period of EoR. However, the
amplitude of this ratio and the sharpness of its transition from
positive to negative is largest at the largest length scales ($k =
0.06\, {\rm Mpc^{-1}}$). At intermediate and smaller length scales ($k
= 0.12\, {\rm and}\, 0.23\,{\rm Mpc^{-1}}$ respectively), its
amplitude reduces significantly and the sharp transition region
becomes more and more flattened.

These features are consistent with the earlier observation of this
quantity using a simulation equivalent to Sem-Num by
\citet{majumdar13}. This sharp peak and dip can be explained by the
contribution from the cross power spectrum term (between density and
neutral fraction or neutral density) in eq. (\ref{eq:lmodel2}) or
(\ref{eq:qlmodel2}). This contribution will be at a maximum in case of
a strictly ``inside-out'' model.  The location and the amplitude of
this feature essentially measures the strength of ``inside-out''-ness
of the simulation. Thus this transition from positive to negative
value can be used as a definite indicator for the onset of
reionization. We observe that all three simulations discussed here
produce these features at the same location (at $\xb \sim 0.8-0.9$)
and with almost the same amplitude (with a maximum of $\sim 10\%$
deviation from each other). Thus one can safely say that the
semi-numerical schemes are robust enough to reproduce the main
observable features introduced by redshift space distortions.

The shaded regions in light and dark gray in the left most panel of Figure
\ref{fig:P2P0_speck} show the uncertainty in the measurement of this
ratio due to the system noise after $2000$ and $5000$ hr of
observation using a LOFAR like instrument. We observe that the
predictions for this ratio by both of the semi-numerical simulations
fall well within the noise uncertainty of LOFAR. At the largest scales
($k = 0.06\, {\rm Mpc^{-1}}$) CPS+GS produces a slightly better match
with \ctwo (deviation is $\le 10\%$ for $0.2 \le \xb \le 1.0$) than
Sem-Num (deviation is $\le 15\%$). Note that at these length scales
uncertainties due to sample variance are expected to be higher than at
smaller length scales. However, at intermediate and smaller length
scales predictions by Sem-Num are closer to \ctwo than those from
CPS+GS are. Note that at smaller length scales the contribution of
noise is expected to be higher but the contribution from sample
variance is expected to be lower.

The next statistical quantity of interest with an observing potential
is the hexadecapole moment ($P^s_4$). If detected, this quantity will
essentially probe the underlying matter density fluctuations. Similar
to the quadrupole moment this can be described through the ratio
$P^s_4(k)/P^s_0(k)$. For a completely neutral IGM both the linear and
quasi-linear model predict a much smaller value for this ratio
($\approx 0.12$) than for $P^s_2(k)/P^s_0(k)$ ($\approx 1.02$). This
will make its detection much more difficult and a longer integration
time or/and higher sensitivity of the instrument would be required. We
find that at the larger length scales relevant for the present day EoR
experiments, the three simulations considered here agree well
(differences $\le 10\%$) in terms of this ratio. However these results
are dominated by sample variance as they fluctuate considerably and
even produce negative values. Therefore we do not include this
quantity in our comparison analysis.

Although not relevant for the comparison presented here, we would like
to note that, all the simulations discussed here, have not taken into
account the effect of spin temperature fluctuations. The assumption of
$T_S \gg T_{\gamma}$ may break down in a situation when 21-cm is
observed in absorption against the CMBR \citep{mao12}. This may happen
at the very early stages of the EoR, when the first astrophysical
sources are formed and they start coupling the spin temperature with
the kinetic temperature of the IGM by Lyman-$\alpha$ pumping. Further,
this period of $T_S<T_{\gamma}$ can be shortend or extended due to the
effect of X-ray heating in the early universe
\citep{mesinger13}. These studies further show that the fluctuations
in the spin temperature can boost the 21-cm powerspectrum by more than
an order of magnitude during this early phase of reionization. The
spin temperature fluctuations may also impact the signal at ten
percent level even when $T_S > T_{\gamma}$ well into reionization. The
fluctuations in the spin temperature due to all these effects may
introduce an additional fluctuation in the 21-cm brightness
temperature $\delta T_b$, which may further complicate the
interpretation of the redshift space 21-cm signal from this era
\citep{mao12,ghara14}. We plan to address these issues in a future
work.

%% file: summary.tex
\section{Summary and Conclusions} 
\label{sec:summary}

A common notion about semi-numerical methods is that they are not
reliable for recreating the ionization history, since they do not
chronologically follow the state of ionization at individual grid
cells. Our comparison between one numerical simulation (\ctwo) and two
semi-numerical simulations (Sem-Num and CPS+GS) does not support this
idea. We find that between Sem-Num and \ctwo the average reionization
history in terms of $\langle \xh1 \rangle_{v}/\langle \xh1
\rangle_{m}$ differs by a maximum of $\sim 5\%$, whereas the same
difference between CPS+GS and \ctwo can be $\sim 10\%$ at the late
stages of EoR. We examine the reconstruction of the reionization
history further by estimating the bias $\left[ b_{z\Delta}(k) \right]$
and cross-correlation $\left[ r_{z\Delta}(k) \right]$ between the
redshift of reionization and density fluctuations at different length
scales. We find that $b_{z\Delta}(k)$ for Sem-Num and CPS+GS is in
excellent agreement ($\leq 5\%$ difference) with that of \ctwo for a
wide range of length scales ($k \leq 1.0 \,{\rm Mpc}^{-1}$). However,
the cross-correlation $r_{z\Delta}(k)$ for CPS+GS is higher than \ctwo
by $\sim 60\%$ (for $k \leq 0.8 \,{\rm Mpc}^{-1}$) and the same for
Sem-Num is higher than \ctwo by $\sim 25\%$.

We have quantified and compared the morphology of the ionization maps
from semi-numerical simulations with that of \ctwo using the bubble
size distribution, the power spectrum and the cross-correlation.  The
bubble size distribution as well as the evolution of $\langle \xh1
\rangle_{v}/\langle \xh1 \rangle_{m}$ reveals that the total volume
ionized in CPS+GS at any stage of the EoR is smaller than in both
Sem-Num and \ctwo. Specifically, CPS+GS produces fewer small
bubbles. The spherically-averaged power spectrum $P_{xx}(k)$, however,
does not show a large difference between the semi-numerical models and
\ctwo. The difference between \ctwo and Sem-Num is within $\sim 15\%$
and the same with CPS+GS is within $\sim 25\%$ for a wide range of
length scales ($0.04 \leq k \leq 2.0 \,{\rm Mpc}^{-1}$) during most of
the EoR ($0.2 \leq \xb \leq 0.85$). The cross-correlation between the
ionization maps of the semi-numerical simulations and that of \ctwo
shows that Sem-Num is strongly correlated with \ctwo ($R_{xx} \ge
0.8$) at large and intermediate length scales ($k \leq 0.7 \,{\rm
  Mpc}^{-1}$; relevant for 21-cm observations), whereas the same for
CPS+GS is relatively poor ($R_{xx} \ge 0.3$). Also, at smaller length
scales the cross-correlation falls more rapidly for CPS+GS than for
Sem-Num. The cross-correlation between the density fields and
ionization maps ($r_{\Delta x}$) shows that the ionization maps in
CPS+GS follows the cosmic web more strongly at all length scales than
the other two schemes. The difference in morphology between the
semi-numerical simulations and \ctwo, especially at small scales, is
likely due to the former's assumption of uniform recombination.

From our analysis of the reionization history, bias $\left[
  b_{z\Delta}(k) \right]$, cross-correlation $\left[ r_{z\Delta}(k)
\right]$, bubble size distribution, power spectrum $\left[ P_{xx}(k)
\right]$ and cross-correlation [$R_{xx}(k)$ and $r_{\Delta x}(k)$], we
can safely conclude that the reionization history and the morphology
of the ionization maps of \ctwo are reproduced with higher accuracy by
Sem-Num than by CPS+GS. These differences are due to the fact that
CPS+GS produces a higher degree of ``inside-out'' reionization, i.e.\
reionization is more biased to denser regions.

The main algorithmic difference between the two semi-numerical schemes
lies in their assumptions regarding the modelling of reionization
sources. Sem-Num takes into account the reionization sources in a
manner very similar to \ctwo. It considers the halos identified from
the $N$-body particle distribution as the hosts of ionizing sources
(eq. [\ref{eq:sem-num}]), whereas CPS+GS does not incorporate halo
masses and locations in its source model, but rather estimates the
collapsed fraction from the density field directly. This causes the
reionization to be more ``inside-out'' in nature for CPS+GS. It also
makes it necessary to smooth the $N$-body density field in CPS+GS
(e.g. by using a Gaussian filter); otherwise both the morphology of
ionized regions and the reionization history become markedly different
from \ctwo.

None of the above quantities are actual observables. The most direct
observable of the reionization process is the redshifted 21-cm signal
from neutral hydrogen.  When comparing the results for this quantity
between the three simulations, we observe that Sem-Num stays highly
correlated (${\mathcal R}_{\delta T_b} \ge 0.8$) with \ctwo at length
scales relevant for present and future experiments such as LOFAR, MWA,
GMRT etc. ($k \leq 0.5 \,{\rm Mpc}^{-1}$) during almost the entire
span of the EoR ($0.2 \leq \xb \leq 1.0$). However, the same
correlation between CPS+GS and \ctwo is much worse (${\mathcal
  R}_{\delta T_b} \ge 0.4 $).

The first observations of the redshifted 21-cm signal will concentrate
on simpler statistical measures, such as the variance of the
signal. We observe that the predictions for the variance from both the
semi-numerical schemes are in well agreement with that of the
\ctwo. The deviation from \ctwo at maximum is approximately $20\%$ and
approximately $30\%$ for Sem-Num and CPS+GS, respectively. These
differences fall well inside the measurement errors of a LOFAR-like
experiment \citep{patil14}.
 
As an aside we find that a correct implementation of redshift space
distortions is important for the 21-cm signal, even when considering
the simplest statistic, namely the variance. The shape and amplitude
of the variance differ considerably between real and redshift space,
especially during the early stages of reionization. Thus it is
essential to incorporate the redshift space distortions accurately
using the actual peculiar velocity fields when trying to constrain the
reionization parameters using the evolution of the redshifted 21-cm
signal from EoR.

We further considered the different angular multipole moments of the
redshifted 21-cm power spectrum. Predictions for the monopole moment
$\left[ P^s_0(k) \right]$ or the spherically averaged power spectrum
from semi-numerical simulations show good agreement with the results
from \ctwo. The $P^s_0(k)$ estimated from Sem-Num and CPS+GS deviates
by $\leq 15\%$ and $\leq 25\%$ respectively from \ctwo at length
scales $k \leq 0.5\, {\rm Mpc^{-1}}$. The power spectrum at these
length scales will possibly become detectable after more than $1000$
hours of LOFAR observations.

The strength and the nature of the redshift space distortions present
in the 21-cm signal can be quantified by the ratio between the
quadrupole and the monopole moments of the redshift space power
spectrum [$P^s_2(k)/P^s_0(k)$] \citep{majumdar13}. The properties and
evolution of this ratio, in principle, can also help in
characterising/constraining the nature of reionization and its
sources. We observe that all three simulations discussed here capture
and represent the major characteristic features of an ``inside-out''
reionization through the ratio $P^s_2(k)/P^s_0(k)$. We find that
Sem-Num performs slightly better (deviation from \ctwo is $\leq 15\%$)
than CPS+GS (deviation from \ctwo is $\leq 20\%$) in terms of the
prediction for this ratio at length scales ($k \leq 0.23\, {\rm
  Mpc^{-1}}$) that will be detectable after more than $2000$ hours of
LOFAR observations. However, both of the semi-numerical results stays
within the noise uncertainties that will be present in such
observations.

In conclusion, we can say that both semi-numerical models discussed
here perform reasonably well in predicting the observables of the
21-cm signal from EoR at length scales detectable with the present and
future experiments, provided that the effect of redshift space
distortions has been implemented in them with an acceptable
accuracy. We also observe that the predictions from Sem-Num are
slightly more similar to \ctwo (by $\sim 10\%$) than CPS+GS for most
of the observables. However, the predictions for the reionization
history and the morphology of the ionization maps are significantly
closer to the benchmark (by $\sim 25 - 50 \%$) in Sem-Num than CPS+GS
mainly due to the differences in their source model. As the
predictions for the 21-cm signal together with the reionization
history would be required for the parameter estimation from the
observational data, it would be better to use a semi-numerical scheme
which can predict both with an acceptable accuracy. We find that among
the two semi-numerical simulations discussed here Sem-Num satisfies
this criterion very well, as it incorporates a source model very
similar to \ctwo. However, we should note that any halo based
simulation technique (radiative transfer or semi-numerical) is
restricted by its particle mass resolution in terms of the dynamic
range that it can explore. On the other hand simulation techniques
based on the conditional Press-Schechter approach is not restricted by
its mass resolution, as the limit on minimum halo mass in the source
model ($\sigma^2(R_{\rm min})$ in the denominator of
eq. [\ref{eq:coll}]) is introduced analytically. This kind of
prescription thus can include atomically or molecularly cooling halos
through this analytical approach.

As we explained in the introduction, we on purpose chose a somewhat
simplified case for our comparison. Here we would like to review
briefly some of the effects which we do not consider.  The first one
is the effect of radiative feedback on the star formation in low mass
halos. Because of the shallowness of their gravitational potential,
halos of masses less than $\sim 10^9 M_{\odot}$ will stop accreting
gas from the IGM once it has been ionized and heated to temperatures
of $\sim 10^4$ K.  This will most likely affect their star formation
efficiency, although the details remain unclear \citep{couchman86,
  rees86, efstathiou92, thoul95, thoul96, gnedin00b,kitayama00,
  dijkstra04, hoeft06, okamoto08}. Since this type of feedback depends
on the distribution of the ionized regions, including it could
increase the differences between numerical and semi-numerical
results. Furthermore, as this effect depends on the local history of
reionization, including it in numerical simulations is rather
straightforward but including it semi-numerical simulations is more
complicated. The reason for this is that in the semi-numerical
approach the photon production in a given region is determined by the
collapsed mass in the chosen mass range of halos rather than on the
actual history of halos in that region. \citet{sobacchi13} recently
proposed a method to include this feedback effect in a semi-numerical
simulation. It would be useful to compare the results of this approach
to a full numerical simulation with a radiative feedback recipe.

The second effect we did not consider is the impact of unresolved
density fluctuations. In fully ionized regions these will boost the
recombination rate and the densest structures will even be
self-shielding for ionizing photons and remain (partly) neutral,
blocking the path of ionizing photons. The effect of these will be to
increase the number of photons needed to reionize the Universe, as a
larger fraction will have to be used for balancing recombinations. The
presence of these structures will also limit the mean free path of
ionizing photons. So-called Lyman Limit Systems are often mentioned in
this context. These features in quasar spectra represent small scale
structures which are optically thick to ionizing photons, although
they do not need to be fully self-shielding. These inhomogeneous
recombinations will limit the growth of ionized regions and therefore
reduce the fluctuations in the 21-cm signal on large scales (e.g. see
\citealt{sobacchi14}, Shukla et al., in preparation). This will
possibly have a significant effect in shaping the spherically averaged
power spectrum of the 21-cm signal from this epoch.

Simulations which do not resolve the full range of density
fluctuations, including the self-shielded systems will thus reionize
too quickly and, due to the long mean free paths for ionizing photons,
produce local photo-ionization rates which are too high. As mentioned
in Section~\ref{sec:RT} for numerical simulations it is possible to
introduce a clumping factor to correct for this lack of resolution
although doing this completely self-consistently is not trivial. The
semi-numerical methods can correct for the enhanced recombination
inside self-shielded neutral regions through sub-grid modelling
\citep{choudhury09,sobacchi14}. However, since these approaches are
not fully equivalent, we have chosen not to include any subgrid
corrections for density fluctuations here.  As for the radiative
feedback, it would be good to compare the different implementations
for the effects of density fluctuations between numerical and
semi-numerical methods. However, we consider this to be beyond the
scope of this paper. We postpone such a comparison to future work.

Lastly, when calculating the 21-cm signal from our simulations, we
have not taken into account the effect of spin temperature
fluctuations. The spin temperature fluctuations due to the
Lyman-$\alpha$ pumping and also due to the heating by X-ray sources
can affect the 21-cm brightness temperature fluctuations significantly
during the early stages of EoR \citep{mesinger13}. These additional
fluctuations in the 21-cm brightness temperature may further
complicate the interpretation of the redshift space 21-cm signal from
this era \citep{mao12,ghara14}. We plan to address this issue in a
future work.